\documentclass[useAMS,usenatbib]{mn2e}

\usepackage{graphicx}
\usepackage{epsfig}
\usepackage{subfigure}
\usepackage{soul}

\newcommand{\be}{\begin{equation}}
\newcommand{\ee}{\end{equation}}

\def \aap{AAP}
\def \apjl{ApJ}
\def \apj{ApJ}
\def \apjs{ApJS}
\def \araa{AAP}

\def \mnras{MNRAS.}

\def \nat{Nature}
\def \pasj{PASJ}
\def \prd{PRD}
\def \aj{AJ}
\def \pasa{PASA}
\def \apss{APSS}
\def\msun{{\,M_\odot}}
\def\lsun{{\,L_\odot}}

\def\lesssim{\lower.5ex\hbox{$\; \buildrel < \over \sim \;$}}
\def\gtrsim{\lower.5ex\hbox{$\; \buildrel > \over \sim \;$}}

\newcommand{\nn}{\nonumber}
\newcommand{\bea}{\begin{eqnarray}}
\newcommand{\eea}{\end{eqnarray}}

\title[Particle Astrophysics of the Galactic Centre]{{\it Wild at Heart:-}\\ The Particle Astrophysics of the Galactic Centre}
\author[Roland M. Crocker, David Jones, Felix Aharonian, et al.]
{R.~M.~Crocker$^{1}$\thanks{E-mail:Roland.Crocker@mpi-hd.mpg.de}\thanks{IIF Marie Curie Fellow},
D. I. Jones$^{1}$, 
F. Aharonian$^{2,1}$, 
C. J. Law$^{3}$,
F. Melia$^{4}$,
T. Oka$^{5}$  
\& J. Ott$^{6}$\\
$^{1}$Max-Planck-Institut f{\" u}r Kernphsik, P.O. Box 103980 Heidelberg, Germany\\
$^{2}$Dublin Institute for Advanced Studies, 31 Fitzwilliam Place, Dublin 2, Ireland\\
$^{3}$Radio Astronomy Lab, University of California, Berkeley, CA 94720, U.S.A.\\
$^{4}$Physics Department, The Applied Math Program, and Steward Observatory, The University of Arizona, Tucson, AZ 85721, U.S.A.\\
$^{5}$Department of Physics, Keio University, 3-14-1 Hiyoshi, Kohoku-ku, Yokohama, Kanagawa 223-8522, Japan\\
$^{6}$National Radio Astronomy Observatory, P.O. Box O 1003, Lopezville Rd, Socorro, NM 87801, USA}
\begin{document}

%%%%%%%%%%%%%%%%%%%%%%%%%%%%%%%%%%%%%%%%%%%%%%%%%%%%%%%%%%%%%%%
%%%%%%%%%%%%%%%%%%%%%%%%%%%%%%%%%%%%%%%%%%%%%%%%%%%%%%%%%%%%%%%

\date{Accepted XXX. Received XXX; in original form XXX}

\pagerange{\pageref{firstpage}--\pageref{lastpage}} \pubyear{2010}

\maketitle

\label{firstpage}

\begin{abstract}

We treat of the high-energy astrophysics of the inner $\sim$200 pc of the Galaxy.
Our modelling of this region shows that the supernovae exploding here every few thousand years inject enough power to
 i) sustain the steady-state, in situ population of cosmic rays (CRs) required to generate the region's non-thermal radio and TeV $\gamma$-ray emission; 
 ii) drive a powerful wind that advects non-thermal particles out of the inner GC; 
iii) supply the low-energy CRs whose Coulombic collisions  sustain the temperature and ionization rate of the anomalously warm, envelope $H_2$ detected throughout the  Central Molecular Zone;
iv) accelerate the primary electrons which  provide the  extended, non-thermal radio emission seen over $\sim$150 pc scales above and below the plane (the Galactic centre lobe); and
v) accelerate the primary protons and heavier ions which, advected to very large scales (up to $\sim$10 kpc),  generate
the recently-identified  WMAP haze and corresponding Fermi haze/bubbles. 
Our modelling bounds the average 
magnetic field amplitude in the inner few degrees of the Galaxy to the range $60 < B/\mu$G $ < 400$ (at 2$\sigma$ confidence) and shows that even TeV CRs likely do not have time to penetrate into the {\it cores} of the region's dense molecular clouds before the wind removes them from the region.
This latter finding
 apparently disfavours scenarios in which CRs -- in this star-burst-like environment -- act to substantially modify the conditions of star-formation.
We speculate that the wind we identify plays a crucial role in advecting low-energy positrons from the Galactic nucleus into the bulge, thereby explaining the extended morphology of the 511 keV line emission.
We present extensive appendices reviewing the environmental conditions in the GC, deriving the star-formation and supernova rates there, and setting out the extensive prior evidence that exists supporting the notion of a fast outflow from the region.
\end{abstract}


%\maketitle

%\renewcommand{\thefootnote}{\arabic{footnote}}

\section{Introduction}
\label{sctn_Intro}

In recent years a picture has emerged that the inner regions  of star-forming galaxies should i) be important sources of $\gamma$-rays in the universe \citep{Pavlidou2002,Thompson2007,Dogiel2009}; ii) drive powerful galactic winds \citep{Dogiel2009} and iii) therefore, be important shapers of the inter-galactic medium, particularly its metallicity \citep{Strickland2009}. 
The Galactic centre (GC) presents our closest view of a Galactic nucleus.  
The physical conditions pertaining in the GC put it on an intermediate footing between the relative quiescence of the Galactic disk interstellar medium (ISM) and the environs at the centre of luminous starbursts.
Still, when viewed from the local perspective, many of the GC ISM parameters are extreme.
The GC proffers, then, perhaps our best hope for a detailed understanding of the processes shaping the evolution of the nuclei of spiral galaxies and, in general, regions energised by galaxy mergers or interactions \citep{Melia2007}. 
Such processes include  star formation in giant molecular cloud complexes, concomitant supernova activity, the resultant generation of nuclear outflows or winds, and the acceleration of non-thermal particle distributions. 

This paper is the third in a series related to the high-energy astrophysics of the Galactic centre (GC).
In the first paper of this series \citep[Paper I]{Crocker2010b} some of us showed that it is likely that 
star-formation and consequent supernova processes in the inner $\sim$200 pc of the Galaxy have, over timescales approaching the age of the Galaxy, 
injected cosmic ray (CR) ions to large heights above and below the Galactic plane creating thereby the very extended regions of diffuse, non-thermal emission recently
uncovered at microwave and $\sim$GeV energies \citep{Finkbeiner2004_II, Dobler2008,Dobler2009,Su2010}.
This scenario requires that the GC has injected an average $10^{39}$ erg/s in CRs over $\gtrsim$5 billion years; 
in this paper we set out independent modelling evidence that supports the notion that the GC supernovae inject, as required, of order this power into non-thermal particles. 

In the second paper of this series  \citep[Paper II]{Crocker2010c} we showed that the GC loses most of the high-energy particles it accelerates -- precisely as required by Paper I.
Specifically, we
determined
the ratios,  $R_\textrm{\tiny{radio}} \equiv  L_\textrm{\tiny{synch}}^\textrm{\tiny{obs}}/L_\textrm{\tiny{synch}}^\textrm{\tiny{thick}}$ and $R_\textrm{\tiny{TeV}} \equiv  L_\textrm{\tiny{TeV}}^\textrm{\tiny{obs}}/L_\textrm{\tiny{TeV}}^\textrm{\tiny{thick}}$, between the GC's {\it observed} non-thermal radio continuum (RC) and  $\sim$TeV $\gamma$-ray emission and that {\it expected} on the basis of the region's FIR output (which traces its current star-formation rate) and the supposition that the region is {\it calorimetric} or a `thick-target' to CRs.
These ratios are
given by $R_\textrm{\tiny{radio}}   \simeq 10^{-1}$ and
$R_\textrm{\tiny{TeV}}  \simeq 10^{-2}$ (both uncertain by a factor $\sim$2).
On the basis of this we found that the GC launches a powerful wind.
This wind serves to advect most of the CR ions and electrons accelerated in the GC from the region, launching them into the Galaxy-at-large.
The identification of this outflow, however, was contingent,  upon GC supernovae acting with typical efficiency as accelerators of CRs (losing about 10\% of the $10^{51}$ erg mechanical energy per SN into CRs);
in this paper we show that, indeed, all the evidence is that GC supernovae do act with at least such typical efficiency.

In this paper we present single-zone, steady-state modelling of the inner Galaxy.
We explain below why both the single-zone and steady-state assumptions are physically plausible though we emphasise that, with respect to the 
GC's TeV $\gamma$-ray emission, others -- notably the HESS discovery paper \citep{Aharonian2006} -- have adopted a non-steady-state interpretation.
Our modelling allows us to render a number of the qualitative conclusions reached in our previous papers more quantitative and show how our modelling of the GC's particle astrophysics allows us to tie down the ranges of a number of parameters including the mean gas density and magnetic field experienced by CRs and the speed of the outflow that advects them from the GC.
Our modelling allows us to delimit the potential contribution made by secondary electrons (and positrons) to the region's non-thermal radio emission and determine the efficiency with which GC SNRs accelerate CR ions and electrons.
We also discuss at some length the consequence of the picture of the GC ISM and its non-thermal particle population that we have developed in the course of the above papers and the modelling described below.

\subsection{Structure of paper}

The remainder of this paper is structured as follows: in \S\ref{sectn_Obsvtns} we delineate the exact region around the GC with which we are concerned and
 describe in detail the RC and $\sim$TeV $\gamma$-ray emission from the GC we seek to understand. 
In \S\ref{modelling} we set out our detailed modelling of the region's high-energy, steady-state particle population and its non-thermal, radiative output.
This affords us an independent handle on the region's ISM parameters and confirms the existence of an outflow as previously suspected.
In \S\ref{section_Results} we describe the results of our modelling and  
in \S\ref{sctn_Implications} we
consider the many implications of our modelling.
These include the determination of
GC environmental parameters (average gas density, magnetic field where CRs are interacting, etc); the result that -- independent of any uncertainty surrounding supernova rate or SNR CR acceleration efficiency -- the GC loses $\sim 10^{39}$ erg/s in non-thermal particles into the Galaxy-at-large; 
and the role this release of CRs plays in explaining large-angular-scale non-thermal emission around the GC.
In \S\ref{sctn_Conclusions} we present our conclusions.
Finally, in extensive appendices we describe
current knowledge about environmental conditions in the GC; set out 
a number of independent determinations of the star-formation and consequent supernova rate in the inner 200 pc; and show that supernovae occur in this region occur on average every $\sim$2500 years.
We also review the many pieces of pre-existing evidence for a powerful GC outflow and set out the expectations for its characteristics as informed by observations of other star-forming galaxies.

\section{Inputs} 
\label{sectn_Obsvtns}

\subsection{Observations of regions of interest}

\subsubsection{TeV $\gamma$-ray data: HESS field defined}
\label{sctn_HESSfield}

The most compact region we concern ourselves with is defined by observations with the H.E.S.S. Imaging Air-Cherenkov $\gamma$-ray Telescope (hereafter, `HESS').
After removal of point TeV sources coincident with Sgr A$^*$ and the SNR G 0.9+0.1 (of angular size $\lesssim 0^\circ.1$), this instrument was shown to have detected \citep{Aharonian2006} diffuse, $\sim$TeV $\gamma$-ray emission distributed more-or-less symmetrically around the GC over the region defined  by $| l | < 0.8^\circ$ and $| b | < 0.3^\circ$ with an intensity of $1.4  \times 10^{-20}$ cm$^{-2}$ eV$^{-1}$ s$^{-1}$ sr$^{-1}$ at 1 TeV (the solid angle denoted by the inner rectangle in Fig.~\ref{fig_2.4GHzmap})\footnote{The HESS field-of-view is $\sim 5^\circ$ and the resolution (FWHM) of the instrument is 11.3$'$.}. 
Only limited and dimmer diffuse TeV emission is detected outside this region. 
Hereafter we denote this region as the `HESS field'. Modelling the particle astrophysical processes that occur within the region seen in projection is the chief motivation of this paper.

Of note is that the spectral index of the GC diffuse $\sim$TeV emission, at $\gamma \ = \ 2.29 \pm 0.07_\textrm{\tiny{stat}} \pm 0.20_\textrm{\tiny{sys}}$
(where the spectrum is defined to be $F_\gamma(E_\gamma) \propto E_\gamma^{-\gamma}$)
 is significantly harder than the spectral index of the CR ion population threading the Galactic disk (i.e., the population we detect at Earth) 
and the HE, diffuse $\gamma$-ray emission they generate.
Disk CRs experience 
energy-dependent confinement (because of the energy-dependence of cosmic-ray scattering on the Galaxy's 
turbulent magnetic field structure) and their resulting steady-state distribution is, therefore, steepened from the 
injection distribution into the softer $\sim E^{-2.75}$ spectrum observed at earth \citep[see, e.g.,][]{Aharonian2006}. 
In fact, the GC TeV $\gamma$-ray spectral index (and that inferred for the parent CR ions) is close to that inferred for the {\it injection} spectrum of Galactic disk CRs, itself within the reasonable range of
$\sim$ 2.1-2.2 expected \citep{Hillas2005} for 1st-order Fermi acceleration at astrophysical shocks (associated, e.g., with supernova remnants). 

Also of significance is the fact that the diffuse TeV $\gamma$-rays and the molecular gas column (as traced by CS line data)
are correlated over the central $\sim 2^\circ$.
This supports the notion that the $\gamma$-rays originate in hadronic collisions between CR ions and ambient hydrogen \citep{Aharonian2006}.
This correlation, however,  breaks down  at Galactic longitude $l = 1^\circ.3$. 
The  HESS collaboration \citep{Aharonian2006} showed that a very plausible explanation of the combined facts of the hardness of the diffuse TeV emission and the breakdown in the correlation at $l = 1^\circ.3$ is that the  emission is a non-steady state phenomenon energised by a single explosive injection of CR ions.
This event should have occurred too recently for detectable diffusion steepening of the CR ion population to have occurred or for the leading edge of the diffusion sphere to have reached  $l = 1^\circ.3$.
Moreover, given the geometry, this event would have to have occurred close to the actual GC and, therefore, plausibly in association with the super-massive black hole Sgr A$^*$.
For a diffusion coefficient typical for that determined in the Galactic plane, the extent of the TeV emission could be used to infer a time since the explosive injection event of $\sim 10^4$ years (this transport timescale being much less that the synchrotron cooling timescale of the $\sim$ 10 TeV electrons required to generate the diffuse TeV emission via IC emission, rendering further support to the notion that the emission is produced by CR ions).

There are steady-state alternatives to the general
scenario described above, however, which we will explore in this paper. 
Generically, these alternatives require that there be an {\it energy-independent} escape 
or loss process acting on the cosmic rays whose characteristic timescale constitutes the smallest relevant timescale 
for the system. 
Given the almost energy-independent $pp$ loss timescale, collisions with ambient gas could satisfy 
this requirement. 
Alternatively, advective transport affected by a GC outflow or wind might
define the shortest relevant timescale. 

\subsubsection{Radio observations of GC diffuse emission: DNS defined}

\begin{figure}
 \epsfig{file=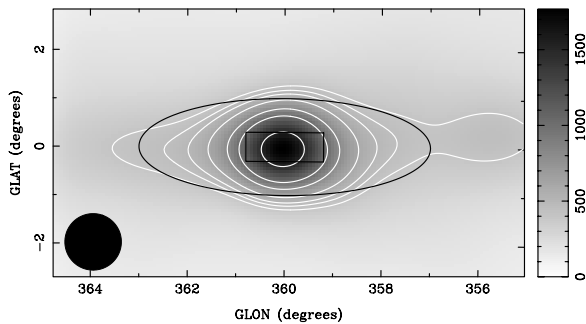,width=\columnwidth}
\caption{Total intensity image
of the region at 2.4 GHz. Radio map \citep{Duncan1995} convolved to a resolution of 1$^\circ.2 \times 1^\circ$.2  with contours at  
400, 450, 500, 600, 800, 1000, and 1500 Jy/beam.
There is a striking constancy in the appearance of the radio structure from 74 MHz to at least 10 GHz.
The large ellipse traces the diffuse, non-thermal radio emission region first identified at 74 and 330 MHz \citep{LaRosa2005}. 
At the assumed distance to the GC of 8 kpc, 1$^\circ$ corresponds to a distance of $\sim$140 pc.
The small  rectangle delineates the HESS field.
}
\label{fig_2.4GHzmap}

\end{figure}

Radio observations at 74 and 330 MHz \citep{LaRosa2005} have revealed a diffuse but discrete, non-thermal radio source (`DNS') approximately covering the  elliptical region of $3^\circ$ semi-major axis (along the Galactic plane) and $1^\circ$ semi-minor axis, centred on the Galactic dynamical centre (corresponding to physical scales of $\sim$ 420~pc and  $\sim140$~pc at the assumed GC distance of 8~kpc); refer Fig.~\ref{fig_2.4GHzmap}. 
This region 
%roughly corresponds to the usual definition of the nuclear bulge of the Galaxy \citep{Launhardt2002,Ferriere2007} and 
easily encompasses the CMZ \citep{Morris1996}, the agglomeration of dense molecular matter along the Galactic plane within $\sim 200$ pc of the actual GC 
%that represents up to 10\% of the Galaxy's $H_2$ allocation (a total mass approaching $10^8 \msun$[\cite{Ferriere2007}]) 
and corresponds, more or less, to the usual definition of the Galaxy's Nuclear 
Bulge \citep{Launhardt2002,Ferriere2007}.
The vertical extension of the DNS (with an approximate half-height of $\sim 150$ pc) corresponds closely to that of the
Galactic centre lobe (see \citet{Sofue1984,Law2010} and references therein) and these structures may be more-or-less identical and, as is investigated below, 
due to an outflow from the central region of the Galaxy.
There is also a suggestive correspondence between the shape,  size, and spectrum of the DNS and the 
extended radio continuum halos seen around many star-forming galaxies, e.g. the
region of 
extended, non-thermal radio emission surrounding the star-burst galaxies NGC 253 \citep{Mohan2005,Zirakashvili2006}
and M82 \citep{Seaquist1991}.
%
%In a detailed analysis (reviewed below) \citet{Law2010} has shown that much of the RC emission from the DNS can be associated with an outflow of GC of material from the GC called the GC lobe \citep{Sofue1984}.

\begin{figure}
 \epsfig{file=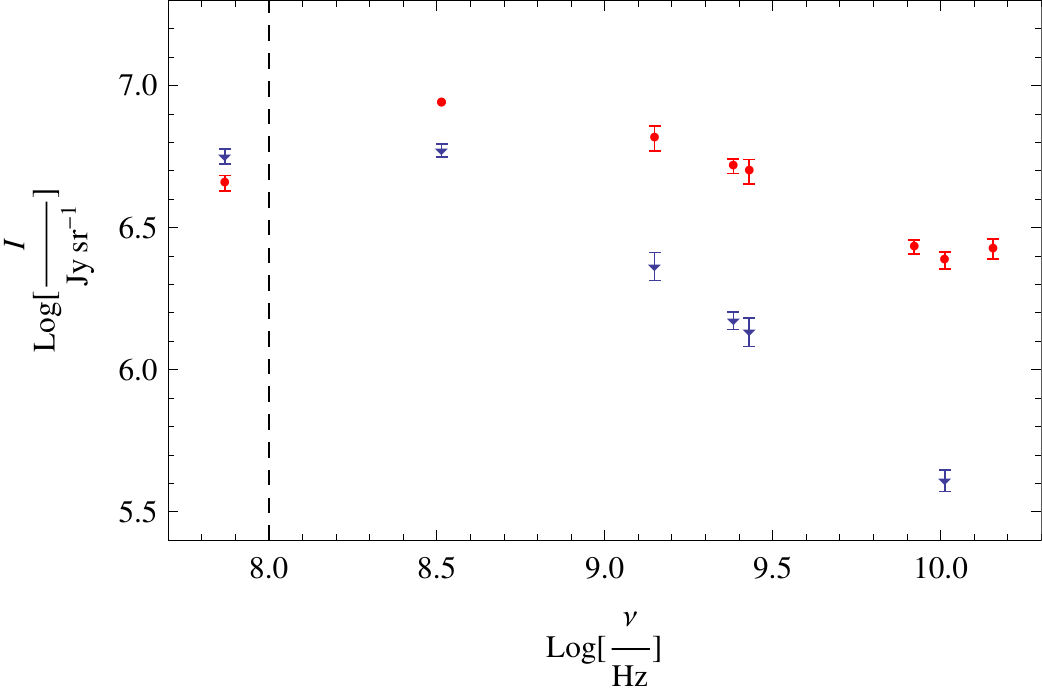,width=\columnwidth}
\caption{Radio spectra for regions under consideration (intensity vs. frequency). 
The triangular (blue) data points refer to the DNS region \citep[see][and references therein and Appendix \ref{sctn_RadioData}]{Crocker2010}. 
The circular (red) data points refer to the HESS field.  
The vertical dashed line divides the (interferometic) 74 MHz data -- which do not receive a contribution from the Galactic plane synchrotron foreground/background -- from all the (single dish) other data (which do).
The 74 MHz data are also strongly affected by free-free absorption.
Note that the 330 MHz flux density for the HESS region \citep{Law2010} is indicative only as it suffers from an uncertain absolute calibration. 
It is therefore {\it not} used in our fitting
described below. 
%The green data points (at 5 and 8.6 GHz) are from observations by \citet{Law2008} of compact and diffuse radio sources within the region centred on the GC and $4^\circ \times 1^\circ$ in extent. These, too, are indicative only and not used in fitting.
}
\label{plotRadioSpctrmPaperII}
\end{figure}

From an examination of archival radio data we have recently shown \citep{Crocker2010} that the spectrum of the DNS is non-thermal up to $\sim$10 GHz with
an overall radio spectral index, $F_\nu \propto \nu^{-\alpha}$ of $\alpha \sim 0.7$ (excluding the 74 MHz datum which is problematic because interferometric and free-free-absorption-affected).
A pure power law is
a poor description of the spectrum (excluded with 3.4$\sigma$ confidence), however, which in fact exhibits
 a spectral steepening at $\sim$1 GHz (see Fig.~\ref {plotRadioSpctrmPaperII}) of $\sim$ 0.6. 
 Our analysis of this radio spectrum -- in concert with $\sim$GeV $\gamma$-ray observations by the EGRET instrument \citep{Hunter1997}  -- has shown that the large-scale magnetic field of the DNS region is at least 50 $\mu$G \citep[radio data is described in detail in the Appendix, the analysis of this data is described at length in the Supplementary Material to][]{Crocker2010}.
Again using archival radio data, we have measured the radio spectrum of the HESS field defined above. This is, as for the DNS, non-thermal up to $\sim$10 GHz (see Fig.~\ref {plotRadioSpctrmPaperII}) with an overall radio spectral index $\alpha \sim 0.5$ (excluding the 74 MHz datum), indicating a very flat synchrotron emitting electron population (and, sub-dominantly, pollution from H{\sc ii} regions as discussed at length below). 
The spectrum is reasonably described by a power law (again excluding the 74 MHz datum).

The total radio flux densities measured within the DNS and HESS fields receive -- in addition to the truly diffuse emission we are concerned with --  contributions from discrete sources within each field and are also polluted by synchrotron emission from CR electrons along the line-of-sight but out of the actual GC. 
How we deal with these back/foregrounds is described at length in section \ref{modelling}  \citep[also see Supplementary Material to][]{Crocker2010}.

\subsubsection{GeV $\gamma$-ray data}

EGRET observations \citep{Hunter1997} show a pedestal of intensity across the inner tens of degrees of the Galactic plane.
No obvious, diffuse source on the same size scales as the DNS radio structure in evident in these data \citep[e.g.,][]{Crocker2010} and the data, therefore, define only upper limits to the GeV intensity that may be predicted by modelling of the region.

\begin{figure}
 \epsfig{file=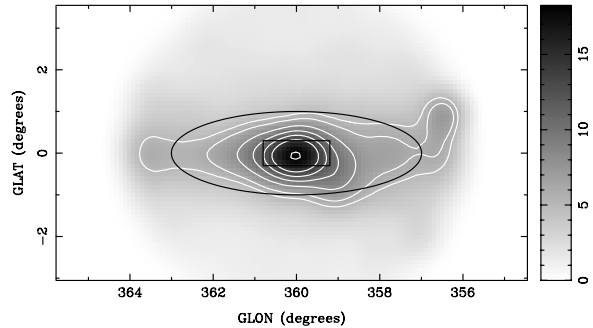,width=\columnwidth}
\caption{Fermi raw count map of  emission above 1 GeV around the GC (as intensity) with contours showing 5 6 8 10 12 14 18 counts (14 months' data-taking) smoothed to the Fermi PSF of 0.3$^\circ$ at 1 GeV illustrating ({\it only}) the $\gamma$-ray morphology. 
The ellipse and box are again the DNS and HESS fields.
The image was made using the standard Fermi data reduction pipeline.
For the {\it total} spectrum of the region we employ the data presented by \citet{Meurer2009}; the GeV point source coincident with $(l,b) = (0,0)$ was recently described by \citet{Chernyakova2010}.
}
\label{plot_FermiMap} 
\end{figure}

In contrast,   it is apparent (see Fig.~\ref{plot_FermiMap}) that the
Femi instrument observes diffuse emission from a region very similar in extent to the DNS.
This $\gamma$-ray emission is, however, again polluted by the contribution of i) individual point sources within the field \citep{Cohen-Tanugi2009} including  a GeV source positionally
coincident with Sgr A and  the GC TeV point source \citep{Chernyakova2010}
and ii) by emission from CRs in the line-of-sight along the Galactic plane but out of the GC \citep{Vitale2009}.
Apparently because of the difficulties attendant upon the removal of these back/foregrounds, the Fermi collaboration has not yet produced an  estimate of the GC's diffuse, GeV $\gamma$-ray emission. 
{\it Total} GeV intensity spectra have appeared \citep{Meurer2009}, however, and {\it these define upper-limits} to what our modelling may predict.
The spectrum of this total emission over the inner GC ($1^\circ \ \times \ 1^\circ$ box centred on the $(l,b)=(0,0)$) is well described by a sharply broken power law that breaks downwards from $\gamma \simeq 1.4$ to $\gamma \simeq 2.6$ at $E_{cut} \simeq 1.6$ GeV \citep{Meurer2009}.
Such a spectrum is consistent with 
 milli-second pulsars (msps) in the GC field (and along the sight) dominating the observed emission. 
The spectrum of individual msps is well-fit as a broken power law with average parameters \citep{Abdo2009b,Malyshev2010} $\gamma = 1.5\pm0.4$, $E_{cut} = (2.8 \pm 1.9)$ GeV and $L = 10^{33.9\pm0.6}$ erg/s.
The number of such pulsars in the GC environment is rather badly constrained given observational difficulties;
broadly, the Fermi GeV data require $\sim$ few hundred msps in the GC field.

\subsubsection{Spatial correlations between GC emission in various wavebands}

Although to be the subject of a forthcoming, in-depth study by some of us\footnote{See Jones, Aharonian, Crocker et al., {\it forthcoming}.}
we mention in passing here that there exists a good spatial correlation between TeV and $\sim$GHz RC emission over the GC \citep[also see][]{Erlykin2007}.
Furthermore, emission in both these wavebands is also correlated with the CS molecular-line data of \citet{Tsuboi1999}, particularly after optical depth effects are taken into account.
These facts support the notion that the observed emission is truly from the GC and that the single-zone assumption we employ (i.e., that CR electrons and protons are generating the observed synchrotron and $\gamma$-ray emission in the same environment) is a justified simplification.

\subsection{Environmental conditions}

Environmental conditions in the GC are reviewed in appendix \ref{sectn_Inputs}.
In general the HESS field encompasses a region of very high volumetric average molecular gas density \citep[$\sim 120$ cm$^{-3}$ as inferred from the data presented by][]{Ferriere2007}, but with most $H_2$ actually concentrated into the low-filling-factor cores of the region's giant molecular clouds.
There is now convincing evidence for a second, relatively low density ($\sim$ 100 cm$^{-3}$) but warmer 
molecular phase enveloping the denser core material.
The dominant filling-factor phase is a very hot plasma of $\sim 10^{(7-8)}$ K \citep[and references therein]{Ferriere2007,Higdon2009}.
The region's magnetic field   is also very strong \citep[$> 50 \ \mu$G:][]{Crocker2010}.
Finally, we note that this paper and Papers I and II reach the conclusion that there is a fast ($\gtrsim$100 km/s) outflow from the GC.
There is considerable {\it prior} evidence for such an outflow which is summarised in appendix \S\ref{sectn_outflow}.

\subsection{Timescales: system in (quasi-) steady state}
\label{section_timescales}

\begin{figure}
 \epsfig{file=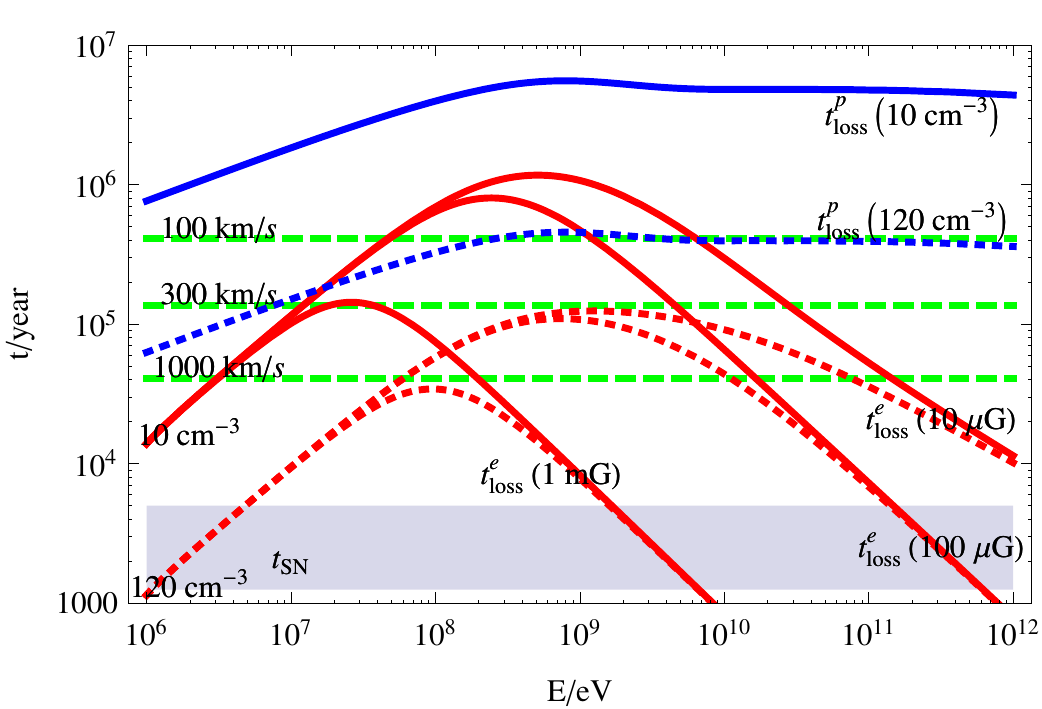,width=\columnwidth}
\caption{ HESS region timescales for i) (horizontal solid band) the inverse of the supernova rate;
ii) (dashed horizontal lines) particle escape (assumed to be advective with energy-independent velocity: lines are, from top to bottom, for cases of $v_{wind} =$ 100, 300, 1000 km/s); 
iii) (red lines) electron cooling (for cases of $B =$ 10, 100, 1000 $\mu$G) in a) $n_H = 10 $ cm$^{-3}$ (solid); and b) volumetric average $n_H \simeq 120 $ cm$^{-3}$ (dotted);
iv) (blue lines) proton cooling in a) $n_H = 10 $ cm$^{-3}$ (solid); and b) volumetric average $n_H \simeq 120 $ cm$^{-3}$ (dotted).
Note that proton energy-loss timescales are always longer than electron energy-loss timescales over the pertinent particle energy range and that calorimetry generically requires $t_{loss} < t_{esc}$.
Particle energy loss processes are described in \S \ref{modelling}.
}
\label{plotTimescales}
\end{figure}

An important consideration in our modelling is whether the system is in steady-state so that, in particular,  the typical time between injection events (i.e., supernovae) -- given the supernova rate -- can be taken to be negligibly small in comparison with other timescales of relevance, in particular, the transport and loss timescales.
In fact, given the supernova rate determined in Appendix \ref{sctn_SFRplusSN}, this condition is met as summarized in Fig.~\ref{plotTimescales} where the various displayed timescales 
can be (approximately\footnote{For full expressions see \citet{Crocker2007}.}) summarized as
\bea
t_\textrm{\tiny{SN}} &\simeq& 2.5 \times 10^{3} \ \textrm{yr} \left( \frac{\nu_\textrm{\tiny{SN}}}{0.04 \ \textrm{(100 \ yr)}^{-1}} \right)^{-1} \ , \nn \\
%t_\textrm{wind} &\equiv& \frac{h}{v_\textrm{wind}}   \simeq 4.1 \times 10^{5} \ \textrm{yr}\  \left(\frac{v_\textrm{wind}}{100 \ \textrm{km/s}}\right)^{-1}  \nn \\
t_\textrm{\tiny{wind}} &\simeq& 4.1 \times 10^{5} \ \textrm{yr}\  \left(\frac{v_\textrm{wind}}{100 \ \textrm{km/s}}\right)^{-1}  \ ,\nn \\
t_\textrm{\tiny{pp}}^p &\simeq& 3.1 \times 10^{5} \ \textrm{yr}\ \left( \frac{n_H}{120 \ \textrm{cm}^{-3}} \right)^{-1}  \ ,\nn \\
t_\textrm{\tiny{inztn}}^e &\simeq& 6.7 \times 10^{5} \ \textrm{yr}\  \left( \frac{E}{\textrm{GeV}} \right) \ \left( \frac{n_H}{120 \ \textrm{cm}^{-3}} \right)^{-1}  \ ,\nn \\
t_\textrm{\tiny{brems}}^e &\simeq& 2.4 \times 10^{5} \ \textrm{yr}\ \left( \frac{n_H}{120 \ \textrm{cm}^{-3}} \right)^{-1}  \ ,\nn \\
t_\textrm{\tiny{synch}}^e &\simeq& 1.3 \times 10^{6} \ \textrm{yr}\ \left( \frac{E}{\textrm{GeV}} \right)^{-1} \left( \frac{B}{100 \ \mu \textrm{G}} \right)^{-2} \ , \nn \\
t_\textrm{\tiny{IC}}^e &\simeq& 1.7 \times 10^{7} \ \textrm{yr}\ \left( \frac{E}{\textrm{GeV}} \right)^{-1} \ .
\eea
(Here the ionization and bremsstrahlung loss time have been adjusted to account for heavier nuclei with $Z>1$ in the background gas.)
Note the hadronic ($pp$) loss time is only slightly longer than the bremsstrahlung loss time with the same direct dependence on $n_H$.
Ionization losses become important for low energy protons but, in the energy range of interest to us, $pp$ is by far the dominant loss process.
The above establishes that the system meets the requirement for being in (at least) quasi-steady state (so that we may seek steady-state solutions of Eq.\ref{keq} below). 

\subsubsection{A note on the TeV $\gamma$-ray deficit at $l = 1^\circ.3$}

As described above, the HESS collaboration interpreted the break-down in the correlation between the diffuse TeV emission and the molecular gas column density as indicating a non-steady-state situation, viz., the $\gamma$-ray emission had been energised by a single CR ion injection event close to the actual GC, the leading edge of which had not yet reached this longitude.
Within the context of the steady-state modelling we perform this explanation is not available to us: this deficit must  speak of either a {\it spatial} change in CR ion population and/or gas characteristics at this distance from the GC.
In fact, such is not unreasonable: 
molecular studies \citep{Tanaka2007} have revealed a distinct
`$l = 1^\circ.3$ complex' which is located (in projection) at the geometrically-privileged position defined by the eastern terminus of the CMZ.
This complex contains gas
characterised by unusually large velocity extents and a `puffy' structure \citep{Oka1996} with at least nine expanding shells,
some possessing enormous ($\sim10^{(51-52)}$ erg) kinetic energy and displaying associated  SiO clumps.
The exact interpretation of this phenomenology is a matter of debate: the structure might have been energised by a very recent burst of massive
star-formation \citep{Tanaka2007}.
Alternatively, the $l = 1^\circ.3$ gas may have only recently arrived in the central region and collided with the central condensation \citep{Sawada2004}.
%so that there has not been enough time for star-formation and subsequent supernova activity to yet run its course in the  $l = 1^\circ.3$ region and this explains the deficit of TeV emission.
%
In any case, given these anomalies, 
a break-down in the correlation between TeV emission and gas density at $l = 1^\circ.3$ -- while certainly interesting -- does not {\it necessarily} reveal
that a single event energised the region's CR ion population; steady-state interpretations of the $\gamma$-ray phenomenology and the particle astrophysics behind it are possible (the simplest probably being that the $l = 1^\circ.3$ gas complex gas is not even contiguous with the rest of the CMZ, but only appears so because of projection effects; we will examine these issues at greater length in future work).

\section{Modelling of Non-Thermal Particle Populations}
\label{modelling}

We have created a one-zone model of the injection, cooling, and escape of relativistic protons, electrons, and secondary electrons (and positrons) from the HESS region.
As justified above, in our modelling we assume that the particle astrophysics can be accurately described to be in quasi-steady state 
and that the particle transport timescale is energy-independent.

\subsection{Modelling}

\subsubsection{Steady-state particle distributions}
We assume homogeneity and isotropy in our one zone modelling. In this case the transport equation is:
\be\label{keq}
{\partial n_x\!(E_x) \over \partial t} = - {\partial \over \partial E_x}
\left[{E_x\, n_x\!(E_x) \over \tau_\textrm{\tiny{loss}}(E_x) }\right]  - {n_x\!(E_x) \over  \tau_\textrm{\tiny{esc}}(E_x)} + 
{\dot Q}_x\!(E_x) \,
\ee
\citep{Ginzburg1964}, where ${\dot Q}_e(E_e)$ denotes 
the injection rate of particles  of type $x$, $x \in \{e,p\}$ and we account for both particle escape and energy loss 
over, respectively, timescales of $\tau_\textrm{\tiny{esc}}$ and $\tau_\textrm{\tiny{loss}}$.
The steady-state distribution of non-thermal particles can be written then as
\bea
\label{solution}
n_x\!(E_x) & = & {\tau_\textrm{\tiny{loss}}\!(E_x) \over E_x} \ \times \ \nn \\ 
&& \int_{E_x}^{\infty} dE_x' \,\, 
{\dot Q}_x\!(E_x') \,\, \exp\!\left[- \int_{E_x}^{E_x'} {dE_x'' \over E_x''} 
\,\, {\tau_\textrm{\tiny{loss}}\!(E_x'') \over \tau_\textrm{\tiny{esc}}\!(E_x'')} \right] \nn \, .
\eea
For our purposes a sufficiently accurate approximation to the above can be written: 
\be
\label{solutionSmpl}
n_x\!(E_x) \simeq \frac{\tau_\textrm{\tiny{loss}}\!(E_x) \tau_\textrm{\tiny{esc}}}{\tau_\textrm{\tiny{loss}}\!(E_x)+(\gamma-1)\tau_\textrm{\tiny{esc}}} {\dot Q}_x\!(E_x)
\ee
where $\gamma$ is the spectral index of the (assumed) power-law (in momentum) proton or electron spectrum at injection, and we assume the escape time, $\tau_\textrm{\tiny{esc}}$, to be energy-independent (see discussion in Appendix \ref{sectn_justification}). 
Here we take it that the acceleration timescale is less than both the loss and escape timescales up to some large $E_x^\textrm{\tiny{max}}$ (otherwise the steady-state cosmic ray proton distribution would not likely be a power law, in contradiction with the expectation supplied by the $\gamma$-ray data).
Note that Eq. \ref{solutionSmpl} reduces to the usual leaky-box expression, $n_x\!(E_x) \simeq \tau_\textrm{\tiny{esc}}  {\dot Q}_x\!(E_x)
$, in the limit that $ \tau_\textrm{\tiny{esc}} \ll  \tau_\textrm{\tiny{loss}}$ and to the `thick target' expression, $n_x\!(E_x) \simeq \tau_\textrm{\tiny{loss}}/(\gamma - 1)  {\dot Q}_x\!(E_x)$, in the opposite limit \citep[see, e.g.,][]{Crocker2007}. 
Given the energy-independent adiabatic energy loss time 
and the
almost energy-independent $pp$ cross-section, Eq.~  \ref{solutionSmpl} renders a very accurate approximation to the steady-state proton distribution. For electrons,
over the parameter space of interest to us (especially for spectral indices of concern which are 2.0 or steeper) we find from numerical evaluation that Eq. \ref{solutionSmpl} reproduces Eq. \ref{solution}
to better than $\sim 35$\%.

\subsubsection{Cooling Processes}

%\subsubsection{Adiabatic cooling}

Relativistic protons and electrons may both suffer adiabatic energy losses in doing $P dV$ work 
on the escaping wind fluid \citep[with the requisite momentum exchange presumably communicated dominantly via the CRs' 
scattering on Alfvenic disturbances in the magnetized wind plasma: e.g.,][]{Socrates2008} which are given by:
\be
 \frac{dE_{CR}}{dt}(E_{CR}) \equiv \, - \frac{1}{3} \nabla . {\bf v} \ E_{CR} \simeq  - \frac{1}{3} t_{esc}  \ E_{CR} \ .
\ee
In addition, as already alluded to, there are cooling processes that affect relativistic electrons or protons exclusively (see \S \ref{section_timescales}).
For protons, note that the steady-state distribution -- given the near energy-independence of the $pp$ cross-section over the apposite energy range (and the assumed energy-independence of the escape time) --will have a spectral shape identical to the injection distribution.
The inelastic collisions suffered by the protons lead to the production of neutral pions (which decay into $\gamma$-rays) and charged pions whose decay products include relativistic electrons and positrons. Self-consistently, our model also tracks the radiative output of these ``secondary electrons".

In the case of electrons (and positrons\footnote{Note that the ultra-relativistic energy scale of the leptons under consideration implies that energy losses for positrons and electrons can be taken to be identical with high accuracy. Positron annihilation can also be ignored: see, e.g., \citet[][]{Crocker2007}.}) we have that
\be
\tau^e_\textrm{\tiny{loss}} (E_e) \equiv - E_e/ \frac{dE_e}{dt}(E_e) \ ,
\ee
where the total loss rate is given by the sum of all loss processes acting to cool the relativistic electron distribution, viz. adiabatic deceleration, ionization, bremsstrahlung, and synchrotron and inverse-Compton emission:
%\beq
 %\frac{dE_e}{dt}(E_e) \equiv  \frac{dE^\textrm{\tiny{inztn}}_e}{dt}(E_e) +  \frac{dE^\textrm{\tiny{brems}}_e}{dt}(E_e) +  \frac{dE^\textrm{\tiny{synch}}_e}{dt}(E_e) +  \frac{dE^\textrm{\tiny{IC}}_e}{dt}(E_e) \ .
%\eeq
\be
 \frac{dE_e}{dt}(E_e) \equiv  \sum_i \frac{dE^\textrm{\tiny{i}}_e}{dt}(E_e) \ ,
\ee
where $ i \in \{ \textrm{inztn},\textrm{brems},\textrm{synch},\textrm{IC} \}$. 
Each of these processes has a characteristic and different energy dependence as described in \S \ref{section_timescales}.
Roughly, the cooling rate is independent of energy in the case of ionization, $\propto E_e$ in the case of bremsstrahlung, and $\propto E_e^2$ in the case of synchrotron and inverse Compton emission: full expressions can be found in \citet{Crocker2007}. 
These different energy dependences introduce spectral features into the steady-state electron distribution that are {\it in addition} to the features inherited from the injection distribution and occur at an energy where one cooling process takes over from another (or from escape).
Note, however, that over the range of energy over which either bremsstrahlung or (energy-independent) escape is the dominant process, the spectral index of the steady-state electron distribution is the same as that at injection.

\subsubsection{Radiative processes}

The same cooling processes described above lead to the radiation of photons that are potentially observable (though it does not follow, of course, that, e.g., for electrons synchrotron radiating at $\sim$GHz frequencies these particles are {\it only} or even dominantly cooled by synchrotron emission). We self-consistently calculate the radiation of the relativistic (primary and secondary) electron population via synchrotron, bremsstrahlung, and inverse Compton emission. We also calculate the $\gamma$-ray emission induced by the hadronic collisions experienced by the relativistic protons. Expression for these processes are given by \citet{Crocker2007}. Note that, in calculating the synchrotron cooling and emission, 
we assume an isotropic electron distribution. This implies the solid-angle-averaged magnetic field perpendicular to the electron's direction of travel is $\langle B_\perp \rangle = \pi/4 \simeq 0.78 \ B$.

\subsubsection{Signal at radio wavelengths}

The (intrinsic, unabsorbed) signal we wish to model at radio wavelengths consists solely of the synchrotron emission due to primary (1e) and secondary electrons (2e):
\be
F_{synch}(\nu) = F^{1e}_{synch}(\nu) \ + \ F^{2e}_{synch}(\nu) 
\ee
where the flux densities of these separate contributions are functions of the following parameters (defined previously) over which we minimize the $\chi^2$ function set out below:
\begin{eqnarray}
F^{1e}_{synch}(\nu) \ = \ F^{1e}_{synch}(\nu , \dot{Q_e}, \gamma_e , t_{esc},B,n_H) \nn \\  
F^{2e}_{synch}(\nu) \ = \ F^{2e}_{synch}(\nu , \dot{Q}_{2e} , t_{esc},B,n_H)
\end{eqnarray}
where $\dot{Q_e}({\rm TeV}) \equiv \kappa_{ep} \dot{Q_p} ({\rm TeV}) $, 
$\dot{Q_{2e}} =  \dot{Q}_{2e}(\dot{Q_p}, \gamma_p , n_H)$, 
 and we assume the injection spectral indices obey $\gamma_e = \gamma_p \equiv \gamma$. 
 Note that, for clarity, here and below we do not list other in-principle parameters that are fixed in our modelling (distance to, volume of, and solid angle of source region).
Also note that at scaling energies
$\gamma_{2e} \simeq \gamma_p$ but we do not {\it assume} this relation. 
Rather, the spectrum of secondary electrons (resulting from the collisions
 of a particular, power-law population of CR protons and taking into account the kinematics of the  $\pi  \ \to \  \mu \ \to \ e$ decay chain)
 is 
as given by a MATHEMATICA interpolation of 
particle yields from accelerator data on charged pion production in $pp$ collisions \cite[as described at length in][]{Crocker2007}\footnote{Data kindly provided by Todor Stanev.}.

\subsubsection{Signal at $\gamma$-ray energies}

At TeV energies we have to consider emission by both (highly) relativistic electrons and protons. Relevant processes are
neutral meson (mostly $\pi^0$) decay following $pp$ collisions for protons and inverse-Compton (IC) and bremsstrahlung for both primary and secondary electrons.
\be
F_{TeV}(E_\gamma) = F^{p}_{TeV} (E_\gamma) \ +F^{1e}_{TeV} (E_\gamma) \ + \ F^{2e}_{TeV} (E_\gamma) 
\ee
where 
\begin{eqnarray}
F^{p}_{TeV} (E_\gamma) &=& F^{p}_{TeV} (E_\gamma , \dot{Q_p}, \gamma , t_{esc},n_H) \nn \\
F^{e}_{TeV} (E_\gamma) &=& F^{IC}_{TeV} (E_\gamma,  \dot{Q_e}, \gamma , t_{esc},B,n_H)  \nn \\
&& \ + \ F^{brems}_{TeV} (E_\gamma,  \dot{Q_e}, \gamma , t_{esc},B,n_H) \ .
\end{eqnarray}
Again, note that the interstellar radiation field --  an important parameter in describing IC cooling and emission -- is fixed in our modelling with an energy density of  $\simeq 90$ eV cm$^{-3}$ and spectrum as described above.
Over reasonable parameter values (in particular, the requisite energy of the emitting electron) synchrotron is not a relevant production process for $\sim$TeV $\gamma$-rays in this environment. 
Also note that we will ignore 
$\gamma \gamma$ attenuation which in principle is non-negligible at energies of $\gtrsim$100 TeV for the photon-rich GC environment \citep{Zhang2006,Moskalenko2006} but which introduces an attenuation which is small on the scale of other uncertainties in our modelling. 
For $pp$ $\gamma$-ray production, $F^{p}_{TeV} (E_\gamma)$, we use an interpolation between (at $E_\gamma \ < \ 100$ GeV) the results of \citet{Kamae2006} and (at $E_\gamma \ > \ 100$ GeV) the results of \citet{Kelner2006} .

\subsection{Back/Foregrounds at radio wavelengths}

As well as the diffuse, synchrotron emission that we seek to determine from our modelling of the region's total RC output, the measured flux density also 
receives  a significant contribution from i) the foreground/background of synchrotron radiation by relativistic electrons in the Galactic plane along the line of sight (but out of the GC), the so-called Galactic Synchrotron Background (GSB; note that the 74 MHz datum is supplied by the The Very Large Array\footnote{The VLA is operated by the National Radio Astronomy Observatory (NRAO). The NRAO os a facility of the National Science Foundation operated under cooperative agreement by Associated Universities, Inc.} an interferometer, 
which is not sensitive to the large-angular-scale GSB at this frequency); ii) discrete radio sources within the field; and iii) free-free (thermal bremsstrahlung) emission from H{\sc ii} regions in the region and in the Galactic plane in front of the GC.

\subsubsection{Galactic plane synchrotron background}

The Galactic plane contribution to the total radio emission is significant -- possibly approaching half the measured emission -- so 
we must deal with this component.
%our modelling procedure is, then, to fit a model of the GC signal plus GSB to the total radio flux observed from the GC. 
%
We normalize our GSB model at 408~MHz via recourse to the  Galactic plane survey of \citet{Haslam1982} and, following
the procedure of \citet{LaRosa2005}, choose a constant longitude slice free of discrete 
sources (well outside of the DNS) near $l = 354^\circ. 5$ to measure the latitude dependence of the 408 MHz flux density normalization.
At higher frequencies we expect the GSB to have a spectral index of $\alpha=-0.70\pm0.12$ \citep{Platania2003} with $I_\nu \propto \nu^\alpha$. Note that the GSB is expected to display a break at a frequency of $\sim 22$ GHz \citep{Voelk1989}, well above the region where we are fitting.
As we cannot directly measure the GSB at higher frequencies due to the presence of other backgrounds, {\it we leave the GSB spectral index as a free parameter in our fitting}. This component of the detected emission is described, then, by a single fitting parameter:
\be
F_{GSB}(\nu,\gamma_{GSB}) \ = \ F_{GSB}(408 \ {\rm MHz})\left(\frac{\nu} {\rm 408 \ MHz}\right)^{-\gamma_{GSB}} \ .
\ee

\subsubsection{Free-free emission and absorption}

Given the compact and extended H{\sc ii} regions within the GC and along the line-of-sight \citep{LaRosa2005} we expect a contribution from free-free (thermal bremsstrahlung) emission to the total radio flux. 
Self-consistently, we must take into account the optical depth due to free-free absorption in modelling the detected radio flux due to synchrotron and free-free emission at the source.
Our modelling of free-free processes works in the following way:
at sufficiently high frequency, the free-free optical depth, $\tau_{ff}$ becomes small and free-free emission, $F^{ff}$, might account for anything up to 100\% of a high-frequency flux datum. Our $\chi^2$ procedure, then, normalizes the total free-free emission at 10.29 GHz where $\tau_{ff} \ll 1$ for reasonable parameters. At this high frequency, the free-free optical depth and emission -- connected through the emission measure -- are related via:
\be
\tau_{ff}(10.29 \ {\rm GHz}) \ = \ \frac{F^{ff} (10.29 \ {\rm GHz})}{\Omega_{\rm HESS} \ B(10.29 \ {\rm GHz},T)} \ ,
\ee
where we make the simplifying assumption that the H{\sc ii} region covers the entire HESS region solid angle\footnote{Note that, in fact, H{\sc ii} absorption over the field is somewhat patchy; \citet{LaRosa2005}.
Though not strictly justified given we find  $\chi^2/dof < 1$ {\it already} for best-fit regions of the current parameter space, we have investigated the introduction of a  further parameter for the H{\sc ii} covering fraction in our modelling.
We find that a covering fraction $\gtrsim 60$\% is required for a good fit; this is consistent with the results of \citet{Mezger1979} who found that most thermal RC emission from the region was contributed by extended, low-density ionized gas (discrete H{\sc ii} regions contributing the remainder).}, $\Omega_{\rm HESS}$, and $B(10.29 \ {\rm GHz},T)$ is the intensity of a black-body of temperature $T$ at 10.29 GHz. 
We adopt $T = 5000$ K in this calculation as motivated by radio recombination line observations of the GC \citep{Ferriere2007,Law2009}.
Self-consistently, the free-free optical depth at any frequency can then be determined as 
\be
\tau_{ff}(\nu) \ = \ \tau_{ff}(10.29 \ {\rm GHz})\left( \frac{\nu}{10.29 \ {\rm GHz}}\right)^{-2.1} \ ,
\ee
and, assuming the geometry of an intervening screen, the {\it detected} free-free emission at any frequency is then
\be
F_{ff}(\nu) \ = \ \Omega_{\rm HESS} \ B(10.29 \ {\rm GHz},T) \ (1 - e^{-\tau_{ff}(\nu)}) \ .
\ee
On the other hand, the synchrotron emission region -- located behind the assumed free-free screen -- has its flux attenuated according to
\be
F_{synch}(\nu) \ = \ F^0_{synch}(\nu) \ e^{-\tau_{ff}(\nu)} \ .
\ee
where  $F^0_{synch}$ is the flux that would be detected from the region in the absence of free-free absorption.

\subsubsection{Discrete sources in the field}

We do not attempt to subtract off the radio flux associated with discrete sources in the field  \citep[which we have previously determined to be at the $\lesssim 10$\% level:][consistent with the expectation that individual SNR contribute 10\% of the total radio flux from normal galaxies: Condon 1992, Zirakashvili \& V{\" o}lk 2006]{Crocker2010} because the TeV $\gamma$-data over the given region does not resolve out any such sources {
\it except} for the Sgr A point source (and the {\it radio} flux associated with the Sgr A source -- in contrast to its TeV flux -- is small in comparison to the region's integrated radio flux; e.g., 20 Jansky at 1.4 GHz, cf. the almost 2000 Jy over the entire region: \citealt{Crocker2007b}). 

\subsection{Modelling procedure}

Our approach is to find -- as a function of environmental and other parameters -- the steady-state populations of relativistic protons and electrons within the region of interest. (Note that in our modelling we neglect for simplicity the poorly-constrained ionic component of the CR hadronic population heavier than protons.) We then self-consistently determine (for the same environmental parameters) the radiative output of these populations, as described above. Finally, we use a $\chi^2$ minimization procedure to determine the parameters describing the proton and electron populations whose radiative outputs in the
given ISM environment
give the best fit
to the 
{\it total} radio flux (density) and the $\sim$TeV $\gamma$-ray flux detected from the HESS field (as previously noted, the GeV data only define upper limits to the broadband emission from the region and we do not attempt to fit them directly).

Environmental parameters that vary within our modelling are magnetic field $B$, ambient hydrogen number density $n_H$ (in whatever form), and the energy-independent timescale over which
particles are advected from the system, $t_{esc} \equiv h/v_{wind}$, where $h \equiv 8$ kpc $\tan(0.3^\circ) \simeq 40$ pc.

We assume that protons and electrons are injected into the GC ISM with distributions governed by power laws in momentum (with identical spectral indices $\gamma_p = \gamma_e = \gamma$). 
The relative normalization of the {\it injection} distribution of electrons to that of protons (at relativistic energies) is given by the coefficient $\kappa_{ep}$ which is also left as a free parameter in our modelling.
The absolute normalization of the distribution of protons at injection is, finally, specified by $\dot{Q_p}$ (in units eV$^{-1}$cm$^{-3}$ s$^{-1}$), also a free parameter within the model

\subsubsection{Chi-squared function}

We determine the volume of parameter space for which an adequate fit to the data is obtained via a numerical minimization 
within MATHEMATICA (using the internal $FindMinimum$ function, with method $InteriorPoint$) of a $\chi^2$ function:
\be
\chi^2 = \chi^2_\textrm{\tiny{radio}} + \chi^2_\textrm{\tiny{TeV}} + \chi^2_\textrm{\tiny{GSB}}
\ee
where 
\be
\chi^2_\textrm{\tiny{radio}} = \sum_i \frac{(F_\textrm{\tiny{radio}}(\nu_i)^{obs} - F_\textrm{\tiny{radio}}(\nu_i))^2}{\sigma^2(\nu_i)}
\ee
with 
\bea
\{\nu_i\} & = & \{74 \ {\rm MHz},1.408 \ {\rm GHz}, 2.417\ {\rm GHz}, \nn \\
&& \ \ \ \ 2.695 \ {\rm GHz}, 8.35 \ {\rm GHz} , 10.29 \ {\rm GHz}\} \nn
\eea
and 
\bea
F_\textrm{\tiny{radio}}(\nu_i) &=& \  \ e^{-\tau_{ff}(\nu)} \left(F^{1e}_{synch}(\nu) + \ F^{2e}_{synch}(\nu) \right) \ \nn \\
&&  \ \ \ \ \ \ \ + \ F_{GSB}(\nu) \ + \ F_{ff}(\nu) 
\eea
and where
\be
\chi^2_{\textrm{\tiny{TeV}}} = \sum_i \frac{(F_{\textrm{\tiny{TeV}}}({E_\gamma}_i)^{obs} - F_{\textrm{\tiny{TeV}}}({E_\gamma}_i))^2}{\sigma^2({E_\gamma}_i)}
\ee
with
\be
F_{\textrm{\tiny{TeV}}}(E_\gamma) = F^{p}_{\textrm{\tiny{TeV}}} (E_\gamma) \ +F^{1e}_{\textrm{\tiny{TeV}}} (E_\gamma) \ + \ F^{2e}_{\textrm{\tiny{TeV}}} (E_\gamma) \ .
\ee
The parameters over which we minimize the $\chi^2$ are $\dot{Q}_p,\kappa_{ep},\gamma,v_{wind},B,n_H,$ and $\gamma_{GSB}$.
%GSB, SPIN, $\kappa_{ep}$, radio and TeV data, total power
%possible constraints: various forms of equipartition, escape speed given by i) Alfven velocity or ii) \citet{Zirakashvili2006} wind velocity, Thompson conditions on $B$ and $n_H$ from $\Sigma_g$
Guided by considerations set out in \S\ref{sectn_outflow}, where necessary to slef-consistently parameterize the wind, 
we assume a thermalization efficiency in the range $0.1 \geq \ \eta \ \geq 0.9$ in our modelling.

\section{Results of Modelling of HESS region}
\label{section_Results}

\subsection{Typical parameters}

Our fitting procedure finds acceptable fits to the data we model for magnetic field amplitudes around
$\sim$ 100 $\mu$G, wind speeds around $v_\textrm{\tiny{wind}} \sim $ few$ \times 100$ km/s, 
total power in all non-thermal particles of $\sim 10^{39}$ erg/s,
gas densities around $n_H \sim 10$ cm$^{-3}$, injection spectral indices $\gamma \sim 2.4$, 
ionization rates of $\zeta \sim 10^{-15}$ s$^{-1}$ and electron to proton ratios (at injection at 1 TeV) of $\kappa_{ep} \sim 10^{-2}$.
(The implications and the statistically acceptable range of each of these are discussed at greater length below.)

\begin{figure}
\epsfig{file= 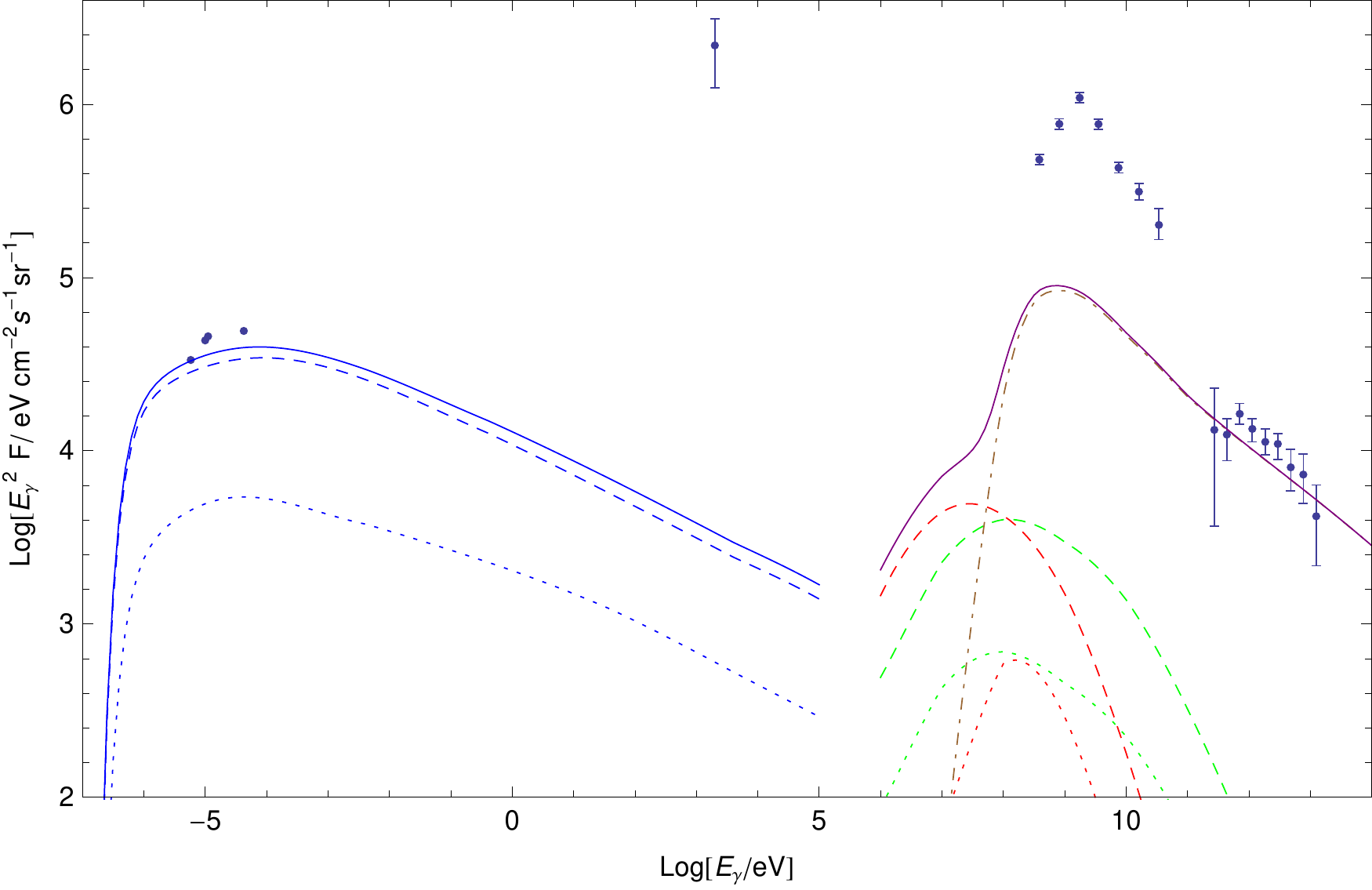,width=\columnwidth}
\caption{HESS field broadband spectrum energy distribution for the best-fit case. Fitted parameters include magnetic field amplitude ($\sim 200 \ \mu$G), ambient gas density ($6$ cm$^{-3}$), at-injection electron-to-proton ratio at 1 TeV ($\sim$0.005), and wind velocity ($\sim$400 km/s). 
Curves are divided into (i) {\bf dashed} -- primary electron emission; (ii) {\bf dotted} -- secondary electron (and positron) emission; and (iii) {\bf solid} -- total emission at a given photon energy. Emission processes are: {\bf blue} -- synchrotron; {\bf red} -- bremsstrahlung; {\bf green} -- inverse Compton; and {\bf brown, dot-dashed} -- neutral meson decay. The total $\gamma$-ray flux is shown in {\bf purple}. 
Data are from (at $E_\gamma \sim 10^{-5}$) radio, 
(at $E_\gamma \sim$ 2 keV) {\it Ginga} \citep{Yamauchi1990} with ,
(at $E_\gamma \sim$ 1 GeV $\equiv 10^9$ eV) Fermi \citep{Meurer2009}, and (at $E_\gamma \sim$ 1 TeV $\equiv 10^{12}$ eV) HESS observations \citep{Aharonian2006}.
We only display modelled synchrotron emission from CR electrons at radio wavelengths; other processes that combine with this to give the region's total radio emission are shown in Fig. \ref{plotRadioSpectrumBestFit} (so the overall fit to the radio data is better than it appears here).
Also note that the Ginga-measured X-ray flux is, apparently, dominantly due to optically-thin thermal plasma emission of temperature $\sim10^8$ K.
The Ginga Large Area Proportion Counter FOV is well-matched to observations on the size scale of the TeV emission and, in fact, show that
the putative thermal plasma covers a very similar solid angle to the HESS region.
Finally, note that the Fermi data points are for total emission from the region and define only {\it upper limits} to the diffuse emission we seek to model.
\label{plotBrdBndSpctraBestFit}}
\end{figure}

\begin{figure}
 \epsfig{file=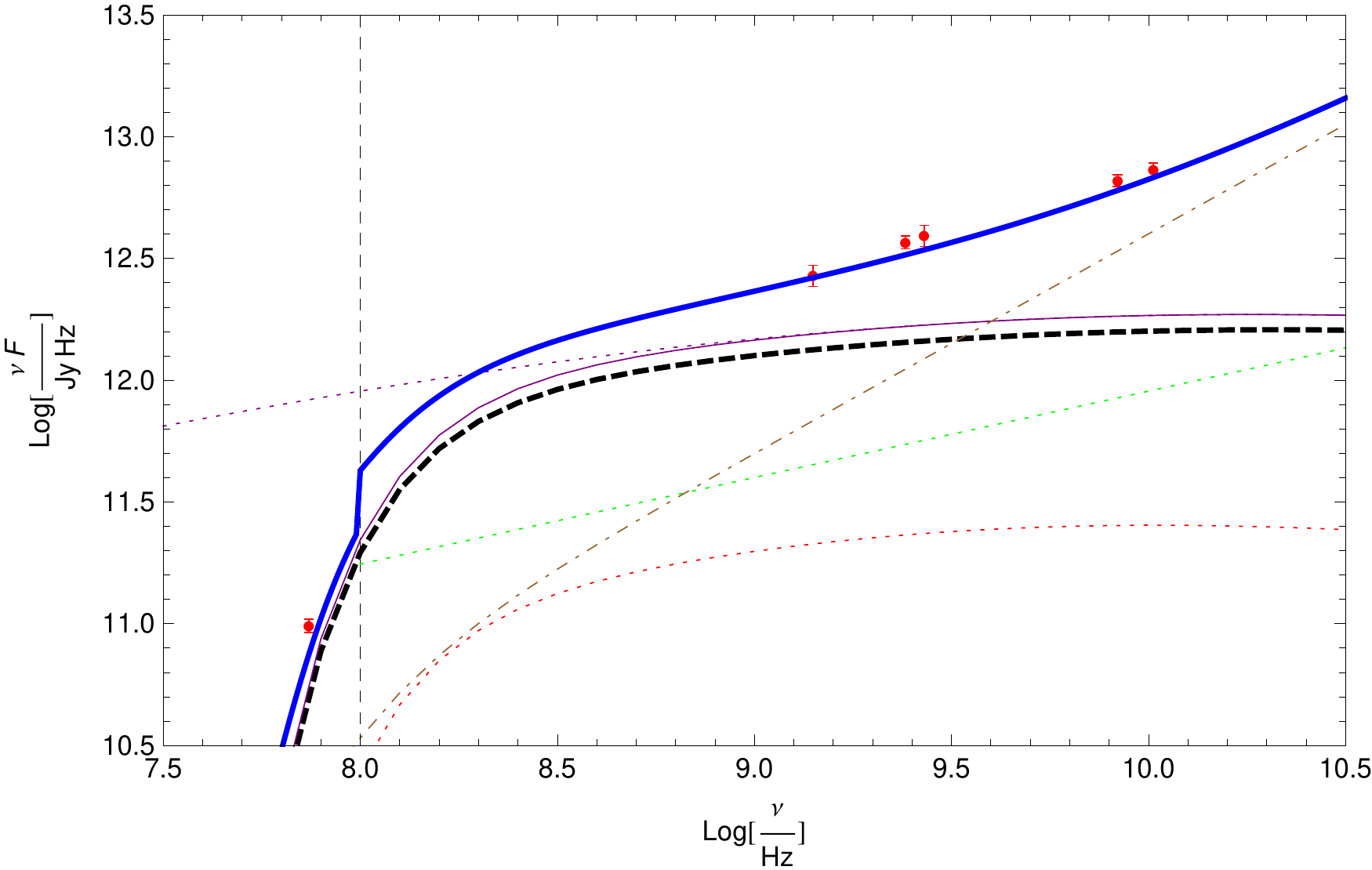,width=\columnwidth}
\caption{Radio data and the modelled radio emission due to various process for the best-fit parameters described in the text. Note that the 74 MHz datum does not receive a contribution from line-of-sight Galactic plane synchrotron (GSB) emission, hence the (unphysical) step in the modelled total emission curve ({\bf solid blue} curve). Other curves are:-- {\bf solid, purple}: total synchrotron; {\bf dotted, purple}: total synchrotron in absence of free-free absorption; {\bf dashed, black}: primary electron synchrotron; {\bf dotted, red}: secondary electron synchrotron; {\bf dotted, green}: GSB; and {\bf dot-dashed, brown}: free-free emission.
\label{plotRadioSpectrumBestFit}}
\end{figure}

\begin{figure}
\subfigure[Modelled proton and electron populations at injection (both assumed to be governed by power laws in momentum) for the best-fit case.]{
 \epsfig{file= 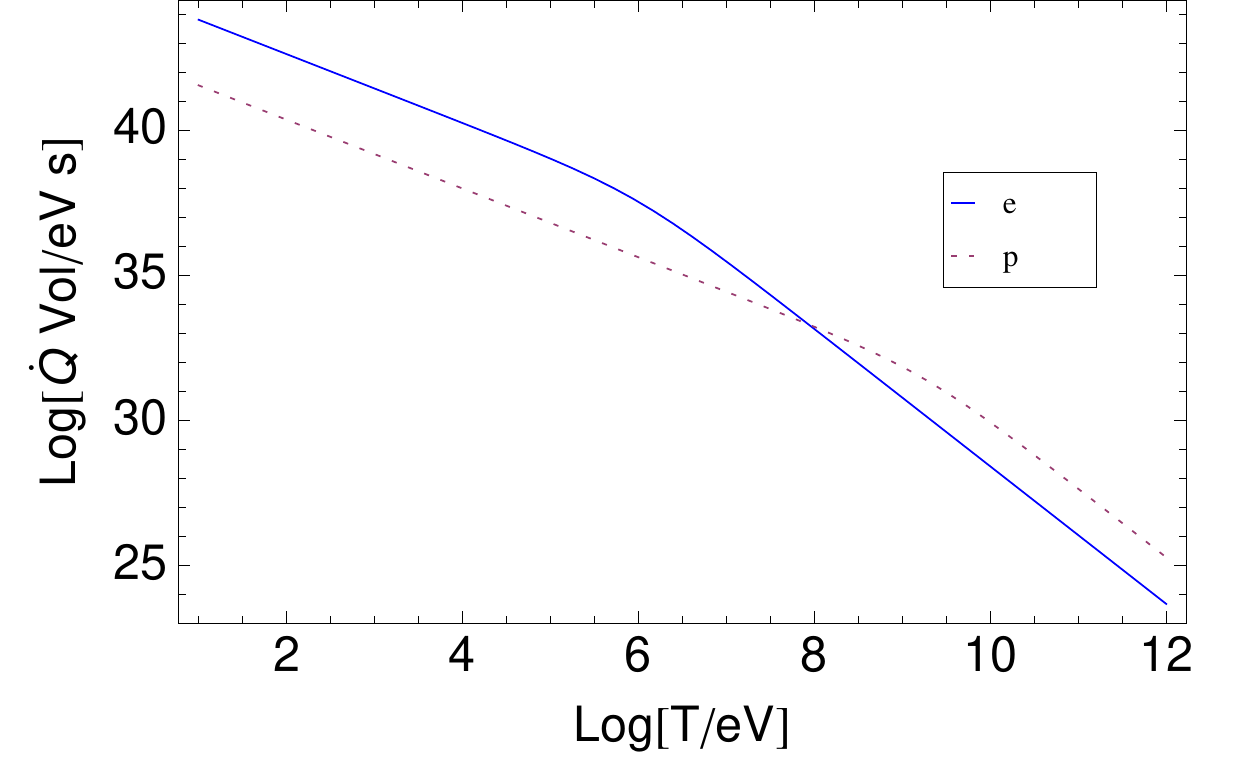,width=0.45\columnwidth}
\label{plotInjctnSpctra}
} 
\subfigure [In situ charged lepton spectra (as formed by the loss and escape processes described in the text) for the best-fit case. The rising spectrum of secondary electrons at very low (and radiatively undetectable) kinetic energies is due to `knock-on' electrons. The local electron spectrum is as described by \citet{Aharonian2009}.]{
 \epsfig{file= 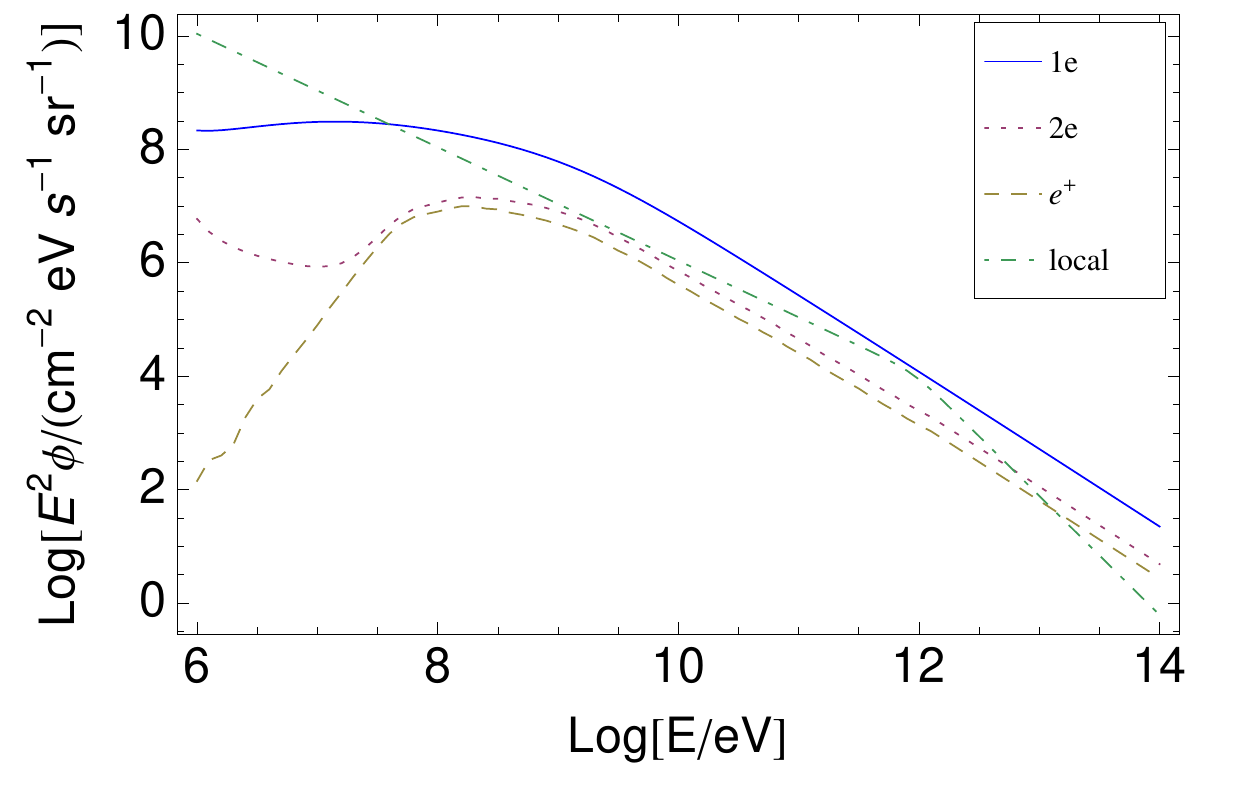,width=0.45\columnwidth}
\label{plotElectronSpectra}
} 
\subfigure [In situ proton spectrum (as formed by the loss and escape processes described in the text) for the best-fit case. The local proton spectrum is as described by \citet{Webber1998}.]{
 \epsfig{file= 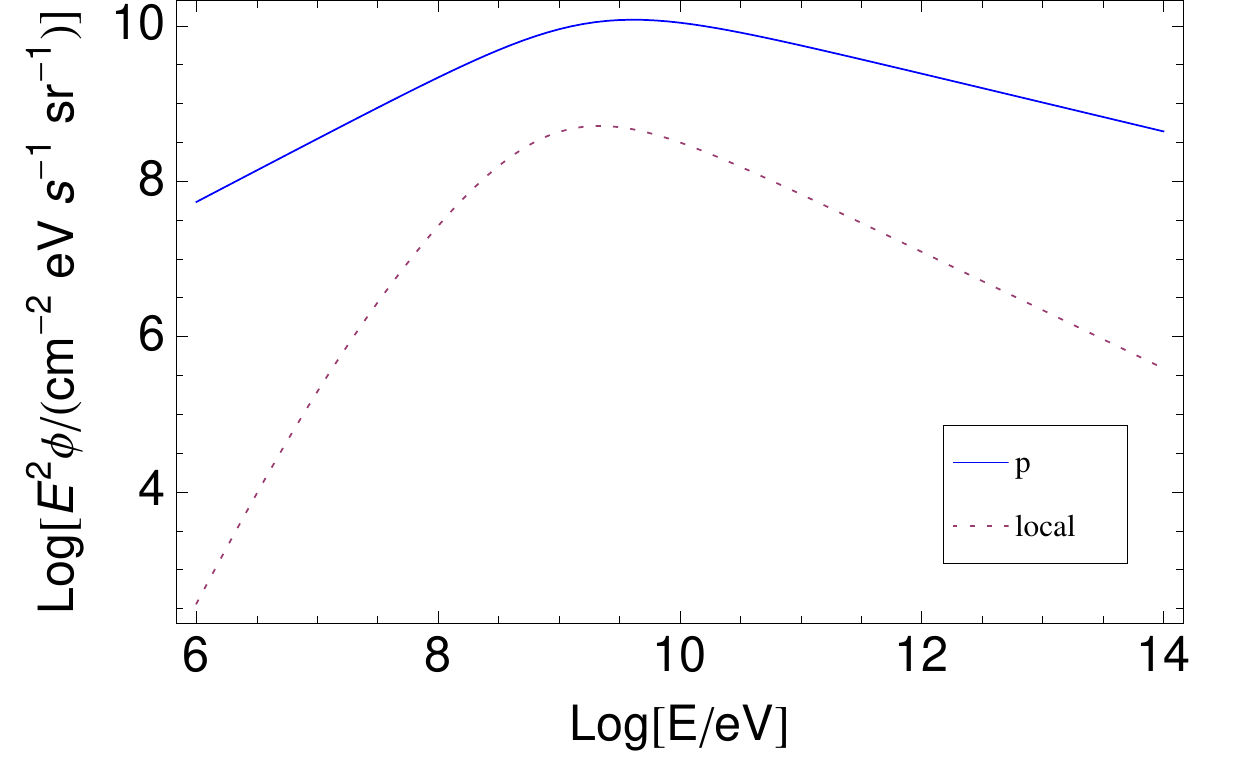,width=0.45\columnwidth}
\label{plotProtonSpectra}
}
%\subfigure [In situ positron fraction (as formed by the loss and escape processes described in the text) for the best-fit case. The data points show the positron fraction detected locally\citep{Adriani2009}.]{
%\includegraphics[width=0.45\columnwidth]{plotPositronFrctn.pdf}
%\label{plotPositronFrctn}
%}
\caption{Inferred particle spectra pertaining to the best-fit case. \label{plotBestFitII}}
\end{figure}

Below we illustrate (figs.~\ref{plotBrdBndSpctraBestFit} and \ref{plotRadioSpectrumBestFit}) the modelled broadband and radio spectra for the best-fit case 
(which achieves $\chi^2 \simeq 7.9$ for $dof = 9$).
We also show (Fig.~\ref{plotBestFitII}) the inferred CR proton and elctron spectra -- at injection and steady-state -- for this particular case.

\subsection{Pure `hadronic', pure `leptonic' scenarios excluded}

As part of our modelling we have examined whether the broad-band, diffuse emission from the HESS region might be explained either within pure `hadronic' or `leptonic' scenarios.
In the former case, synchrotron emission by secondary electrons and positrons, by construction, explains most of the non-thermal RC emission while the non-thermal TeV emission is mostly accounted for by neutral pion decay (with a subdominant contribution from secondary electron IC and bremsstrahlung). Note that \citet{Thompson2006} have claimed an important -- even dominant -- role for secondary electrons in generating the RC detected from star-bursting systems.
In a `leptonic scenario', primary electron synchrotron explains the RC while
 IC and bremsstrahlung by higher-energy members of the same primary population explains the TeV signal.
In fact we find poor fits for both these limiting cases: the best-fit pure hadronic case has $\chi^2/dof = 14.9/7$, acceptable only at $\sim 2.1\sigma$ and requires a very strong ($\sim$4 mG) field;  the best-fit pure leptonic case has $\chi^2/dof = 28.3/7$, acceptable only at $\sim 3.7\sigma$ and requires a very weak ($\sim$16 $\mu$G) field and a very tenuous ($\sim2$ cm$^{-3}$) gas environment.
%

%\begin{figure}
%{\includegraphics[width=\columnwidth]{tableExampleFit.pdf}
%\caption{Various parameters as determined for the best-fit case. }
%\label{table_ExampleFit}
%} 
%\end{figure}

\subsection{Existence of nuclear wind confirmed by modelling}

\begin{figure}
 \epsfig{file=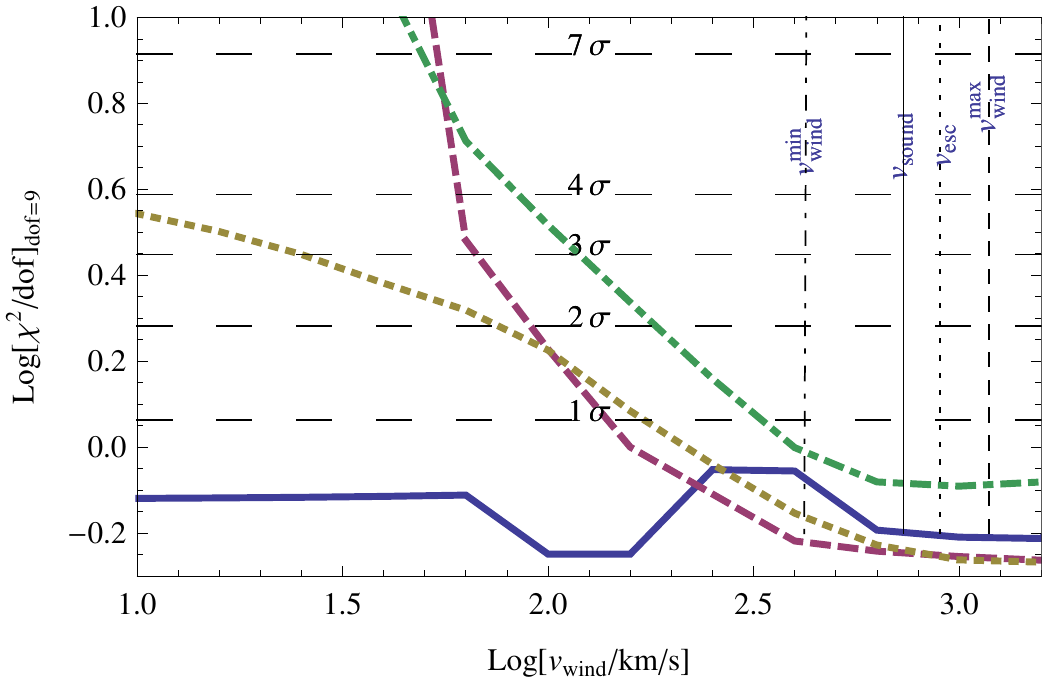,width=\columnwidth}
\caption{~$\chi^2/dof$ as a function of $v_{wind}$. Solid curve is for a $\chi^2$ obtained from a pure fit to the broadband data as described in the text ($dof$ = 9).
The dashed (purple) curve follows the
introduction of a term into the $\chi^2$ for power injected into CRs (central datum value set to power expected on basis of SN rate plus stellar winds, $1.4 \times 10^{39}$ erg/s) and constrains the power advected from the system by CRs to be enough to generate the DNS emission, $\dot{E}_{adv}> 10^{38}$ erg/s
($dof$ = 10).
The dotted (yellow) curve follows the imposition of the heuristic \citep[e.g.,][]{Voelk1990}  constraint that the ratio of proton to electron differential number density injected into the ISM satisfy $\kappa_{ep} < 0.01$ and also that $n_H \geq n_{H^+} \equiv 6$ cm$^{-3}$, where $n_{H^+}$ is the volumetric average 
over the HESS region
of {\it all} gas in plasma phase as inferred from the data of \citet{Ferriere2007}.
The dot-dash (green) curve shows a fit to the broadband data which also attempts to reproduce $R_\textrm{\tiny{TeV}} \simeq 0.01$ and $R_\textrm{\tiny{radio}} \simeq 0.1$ (see Paper II and \S\ref{sctn_Intro}.) with 
 $n_H > n_{H^+}$ = 6 cm$^{-3}$.
The vertical dashed and dot-dashed lines show, respectively the approximate maximum ($\sim1200$ km/s) and minimum wind speeds according to Eq.~\ref{eqn_ZV_vel} with thermalization efficiencies of $\eta_\textrm{\tiny{therm}}^\textrm{\tiny{max}} \equiv 1$ and $\eta_\textrm{\tiny{therm}}^\textrm{\tiny{min}} \equiv 0.1$.
The vertical dotted line shows the approximate gravitational escape speed from the inner GC of $\sim$ 900 km/s \citep{Muno2004} and
the vertical solid line shows the mid-plane sound speed, $c_s = \sqrt{\gamma \ P_c/\rho_c}$, in the wind plasma for central values of $\dot{E}$ and $\dot{M}$.
}
\label{plotChiSqrdvWind}
\end{figure}

Pure modelling of the non-thermal spectrum does not constrain the wind speed but the imposition of physically-motivated constraints (as described in the Fig.~\ref{plotChiSqrdvWind} caption) on the parameter space shows that the wind speed is $> 200$ km/s and most likely several hundred km/s (see Fig.~\ref{plotChiSqrdvWind}).
Given the $1.4 \times 10^{40}$ erg/s mechanical power supplied by supernovae, a conservative (because assuming 100\% thermalization and no mass loading) upper limit on the wind speed is $\sim 1200$ km/s.

\subsection{GC magnetic field determined by modelling}

\begin{figure}
 \epsfig{file=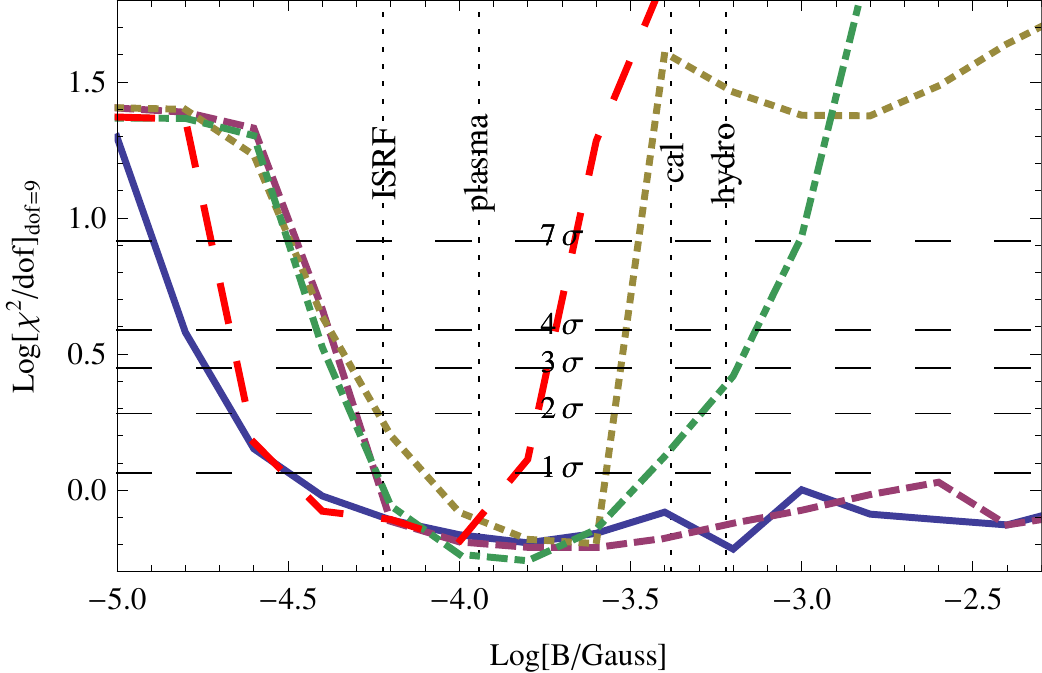,width=\columnwidth}
\caption{~$\chi^2$ as a function of $B$. 
The {\bf solid curve} is for a $\chi^2$ obtained from a pure fit to the broadband data as described in the text ($dof$ = 9) with wind speed as an independent variable.
The {\bf short dashed curve} is a pure fit to the broadband data as described in the text ($dof$ = 9) with wind speed as an independent variable and $\kappa_{ep} < 0.01$.
The {\bf dotted   curve} requires the wind speed obey Eq.~\ref{eqn_ZV_vel} with $\eta_{\rm therm}$ a free parameter (constrained to lie in the range $0.1 < \eta_{\rm therm} < 0.9$)  with the constraint that the total non-thermal pressure (generated by the modelled magnetic field and CR population) be no larger than the pressure launching the wind as given by Eq.~\ref{eqn_Pc} ($dof$ = 9).
The {\bf dot-dashed curve} also requires the wind speed obey Eq.~\ref{eqn_ZV_vel} with $\eta_{\rm therm}$ a free parameter and, in addition, 
introduces a term into the $\chi^2$ for power injected into CRs (central datum value set to power expected on basis of SN rate plus stellar winds, $1.4 \times 10^{39}$ erg/s) and constrains the power advected from the system by CRs to be enough to generate the DNS emission, $\dot{E}_{adv}> 10^{38}$ erg/s
($dof$ = 10).
The {\bf long dashed curve} constrains the magnetic field and CRs to be in energy-density equiparition under the assumption that 
both these phases are volume-filling.
The vertical dotted lines show the magnetic field amplitude 
i) in equipartition with the interstellar radiation field (`ISRF'), 
ii) in equipartition with the very hot plasma (`plasma'), 
iii) required for electron calorimetry (defined by $t_{cool}^e \equiv t_{esc}$ and conservatively assuming $n_H = \langle n_H \rangle_{vol} \simeq 120$ cm$^{-3}$, $v_{wind}^{min} \simeq 760$ km/s as supplied by Eq.~\ref{eqn_ZV_vel} and adopting $\eta_\textrm{\tiny{therm}}^{min} \equiv 0.1$),
and iv) required to achieve $by itself$ hydrostatic equilibrium.
}
\label{plotChiSqrdB}
\end{figure}

As remarked upon at length in Appendix \ref{sctn_GCBFld}, the GC magnetic field has been uncertain by about two orders of magnitude for some time though our recent work \citep{Crocker2010} has determined a lower limit to the field on the size scale of the DNS of 50 $\mu$G.
Relaxing all constraints and fitting only to the broadband spectrum we do not significantly constrain this range.
If we impose the reasonable \citep[e.g.,][]{Voelk1990} -- but admittedly heuristic -- constraint that the ratio of proton to electron differential number density injected into the ISM satisfy $\kappa_{ep} < 0.01$, we find that the magnetic field is required to satisfy $B \ > \ 50 \ \mu$G   at 2$\sigma$ confidence with a minimum at $\sim$160 $\mu$G.
Requiring that the outflow velocity is governed by  Eq.~\ref{eqn_ZV_vel} in addition to other, 
reasonable constraints (see caption of Fig.\ref{plotChiSqrdB}), tends to pick out best-fit $B$ values around this value. 
It is interesting to remark, moreover, 
that imposition of these extra constraints requires the field be $\lesssim 300 \ \mu$G, less than the
 $\sim$600 $\mu$G necessary for the magnetic field {\it alone} to ensure hydrostatic equilibrium (see Appendix \ref{section_Bfield}). 

This seems to tell against a \citet{Thompson2006}-type scenario, at least as it might apply to the GC: the magnetic field in the GC -- while certainly much larger than that in the disk -- is neither large enough to ensure the region is calorimetric to electrons nor (alone) to establish hydrostatic equilibrium
(though one should keep in mind that the region is not in hydrostatic equilibrium in any case given the evidence for an outflow which is driven by the {\it combined} pressures of the plasma and non-thermal components).
If the situation really is as described by \citet{Thompson2006} in dense star-bursts, this is, then, a point of difference between the GC and these systems.

The field is also significantly weaker than the $\sim$ mG amplitude suggested by analysis \citep {Yusef-Zadeh1987,Morris1989} of the region's non-thermal filaments (NTFs; see Appendix \ref{sctn_GCBFld}).

\subsection{Power demanded by modelling}

\begin{figure}
 \epsfig{file= 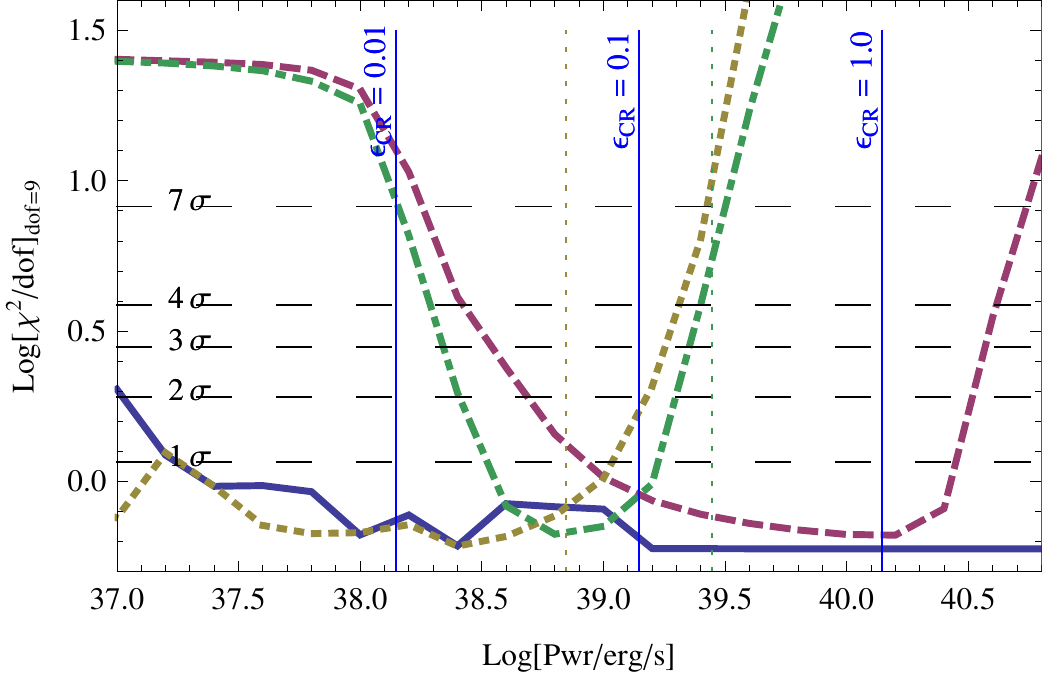,width=\columnwidth}
\caption{~$\chi^2$ as a function of power injected into all CRs and CR electrons. 
The solid curve shows a fit to the broadband data only ($dof = 9$) for total CR power.
The dotted curve shows a fit to the broadband data only ($dof = 9$) for total CR electron power.
The solid vertical lines show the power in non-thermal particles assuming they capture a factor
$\eta_{CR}$  of 0.01, 0.1, and 1 (as marked) of the total mechanical power released into the HESS region
by supernovae (and stellar winds), adopting the central value of supernova rate of 0.04/century.
(The vertical dashed lines indicate, for $\eta_{CR} = 0.1$ the range of  the power given by the factor of $\sim$ 2 uncertainty in the supernova rate.)
The dashed curve shows a fit to the total CR power constrained such that the wind speed obey Eq.~\ref{eqn_ZV_vel} with $\eta_{\rm therm}$ a free parameter (constrained to lie in the range $0.1 < \eta_{\rm therm} < 0.9$)  with the constraint that the total non-thermal pressure (generated by the modelled magnetic field and CR population) be no larger than the pressure launching the wind as given by Eq.~\ref{eqn_Pc} ($dof$ = 9).
The dot-dash curve shows the equivalent constraints for the case of CR electron power.
}
\label{plotChiSqrdPower}
\end{figure}

From direct modelling of the broadband spectrum we find 
no clear minimum in the $\chi^2$,
expressed as a function of power injected into all CRs; see Fig. \ref{plotChiSqrdPower}.
This figure, however, also shows the effect of introducing
the dual constraints of i)
requiring that the  pressure contributed by CRs and magnetic fields (Eq.~\ref{eqn_Pc}) is no larger than the total pressure launching the wind
and ii) self-consistently setting the 
wind speed to that given by the mechanical power and mass injected into the system (while still allowing the thermalization efficiency to vary between 0.1 and 0.9 -- see Eq.~\ref{eqn_ZV_vel}).
With the imposition of  these constraints the allowed parameter space is significantly reduced, favouring a power into non-thermal particles $\gtrsim 10^{39}$ erg/s, the value predicted by the supernova rate and adopting a fiducial $10^{50}$ erg per SN into CRs.
In other words, we reach the interesting conclusion that GC SNRs are at least as efficient as disk SNRs as CR accelerators, losing 
$\gtrsim$10\% of their mechanical power into acceleration of such particles.

\subsection{Medium where emission is occurring}
\label{sctn_Medium}

\begin{figure}
 \epsfig{file=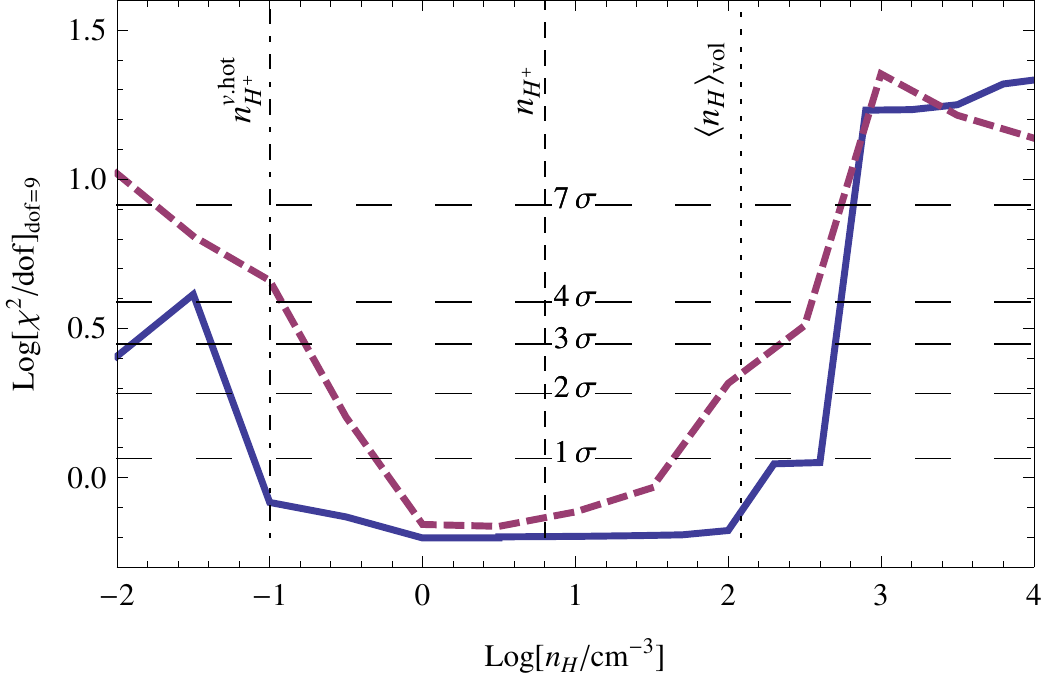,width=\columnwidth}
\caption {~$\chi^2/dof$ as a function of $n_H$ (strictly for $dof = 9$). 
Solid curve is for a $\chi^2$ obtained from a pure fit to the broadband data as described in the text ($dof$ = 9) with wind speed as an independent variable.
The dashed curve fits to the broadband data and also attempts to reproduce $R_\textrm{\tiny{TeV}} \simeq 0.01$ and $R_\textrm{\tiny{radio}} \simeq 0.1$ (see Paper II and \S\ref{sctn_Intro}) with wind speed given by Eq.~\ref{eqn_ZV_vel} with $0.1 \leq \eta_{\rm therm} \leq 1.0$ ($dof$ = 10). 
The vertical dashed line shows the $n_{H^+}$, the volumetric average $n_H$ over the HESS region due to gas in all plasma phases.
The vertical dot-dashed line shows the density, $n_{\tiny{H^+}}^{\tiny \textrm{v.hot}} \sim 0.1$ cm$^{-3}$ \citep{Muno2004}, of {\it only} the very hot phase plasma phase
The vertical dotted line shows the volumetric average $n_H$ over the region for all matter as inferred from \citet{Ferriere2007}.
}
\label{plotChiSqrdnH}
\end{figure}

On the basis of our modelling we can infer the characteristic gas density for the region in which the modelled non-thermal radiation is generated.
Fitting only to the broadband data we find a very shallow minimum $\chi^2$ for 1 cm$^{-3} \gtrsim$ nH $\gtrsim 100^{-3}$ cm$^{-3}$ with an even larger $n_H$ range acceptable at $2\sigma$ level (see Fig. \ref{plotChiSqrdnH}). Note that the volumetric average $n_H$ in the region is $\sim 120$ cm$^{-3}$ with most of the mass contributed by molecular gas found at very high number density ($\gtrsim 10^4$ cm$^{-3}$) in the cores of the region's giant molecular clouds.
It is clear that the hypothesis that emission originates dominantly in these regions is disfavoured.
We also indicate in the figure the volumetric-average number density of gas in all plasma phases,  $n_{H^+}$, \citep[including the `very hot', `hot', and, dominantly, `warm' H$^+$ phases identified by][see Appendix \ref{sctn_Plasma}]{Ferriere2007}.
Given that, according to the analysis of \citet{Higdon2009}, relativistic particles will not be trapped in the very hot phase (for which $n_H \leq 0.1$ cm$^{-3}$)
$n_{H^+}$ arguably constitutes a lower limit to the effective, path-averaged $n_H$ the region presents to CRs.

Inclusion of data in addition to the broadband spectrum constrains the allowed $n_H$ parameter space in a very interesting fashion. The dashed and dotted curves in Fig. \ref{plotChiSqrdnH}  shows the effect of including terms for $R_\textrm{\tiny{TeV}} (\simeq 0.01$) and $R_\textrm{\tiny{radio}} (\simeq 0.1$) in the $\chi^2$ (see Paper II and \S\ref{sctn_Intro}) and 
replacing the previously-independent wind speed parameter with Eq.~\ref{eqn_ZV_vel} with $\eta_\textrm{\tiny{therm}}^\textrm{\tiny{min}} = 0.4$ and $\eta_\textrm{\tiny{therm}}^\textrm{\tiny{max}} = 0.9$, respectively ($dof$ = 11) with the maximum thermalization scenario favouring higher densities and the minimum lower.
Over $\eta_{\rm therm}^{min} < \eta_{\rm therm} < \eta_{\rm therm}^{max}$ the 2$\sigma$-acceptable region now extends from 0.5 cm$^{-3} < n_H < 80$ cm$^{-3}$, with even the volumetric $n_H$ value formally excluded at the 2$\sigma$ level.
If we accept $n_{H^+}$ as a plausible lower limit,  we find an $n_H$ in the range 6-80 cm$^{-3}$ is favoured by our analysis. 
This -- together with the best-fit magnetic field described above -- is consistent with the region of parameter space favoured by the
heuristic arguments presented in Paper II.

This result hints that CRs -- even those at super-TeV energies  -- do not penetrate freely into dense molecular cloud material; cf. theoretical  \citep{Dogiel1990,Gabici2007} and empirical studies \citep{Protheroe2008,Fujita2009} of this issue\footnote{Also see Jones, Crocker et al. 2010, `Australia Telescope Compact Array Radio Continuum 1384 and 2368 MHz Observations of Sgr B', submitted to the Astronomical Journal.}. 
Given the observational fact of the angular correlation between both RC and TeV $\gamma$-ray emission and the molecular material (as traced by CS),
these results are consistent with the observed radiation being generated in low density H{\sc i} or H{\sc ii} material surrounding the clouds \citep[cf.][]{Higdon2009} or, indeed, their relatively tenuous and warm molecular envelopes of the cores (but not dominantly in the hottest, lowest-density, volume-filling H{\sc ii} phase: cf.~Fig.~\ref{plotChiSqrdnH}).

These results are also potentially consistent with recent findings from Fermi observations of the LMC \citep{Knodlseder2009}. 
Here it was demonstrated that neutral and molecular hydrogen templates poorly fit the GeV data whereas an ionized hydrogen template (tracing current star formation) provided a relatively 
good fit. 
In the GC, our analysis suggests that the synchrotron emission from $\sim$GeV electrons and $\gamma$-ray emission from super-TeV protons may mirror this result.
In part, this can be simply explained as a result of the fast advection timescale: the non-thermal particles are advected from the system in a timescale $\sim 4 \times 10^4 \ v_{\rm wind}/$(1000 km/s) yr to be contrasted with the advection timescale {\it into} the giant molecular cloud complexes of 
$t_{\rm adv} \equiv	R_{\rm cloud}/\sigma_v$	which for, e.g., Sgr B ($R_{\rm cloud} =12$ pc and $\sigma_v = 40$ km/s) evaluates to $t_{\rm adv} = 3 \times 10^5$ yr
\citep[of course, this comparison neglects the poorly constrained diffusion timescale into the clouds;][]{Protheroe2008}

We note in passing that, as stated above, our modelling makes the simplifying assumption of a single zone.
This implies, in particular, that the 
synchrotron emission we model -- due to $\sim$ GeV energy electrons -- can be treated as though it is 
emanating from the same environment as the
TeV signal which is  (predominantly) generated by impacts of multi-TeV protons on ambient gas.
There is the possibility, of course, that this assumption does not accurately reflect the circumstances in the GC; that, in fact, multi-TeV protons sample, on average, denser gas than the CR electrons.
We have investigated this possibility in modelling which allows for different environmental parameters ($B$ and $n_H$) in describing the CR electron and proton environments. 
This modelling actually prefers, if anything, electron environments slightly denser than the proton ones.
We take this as a good indication that the single-zone modelling described above (fortuitously or not) renders a good description of the GC's high-energy astrophysics.

\subsection{Injection ratio of electrons to protons}

\begin{figure}
 \epsfig{file= 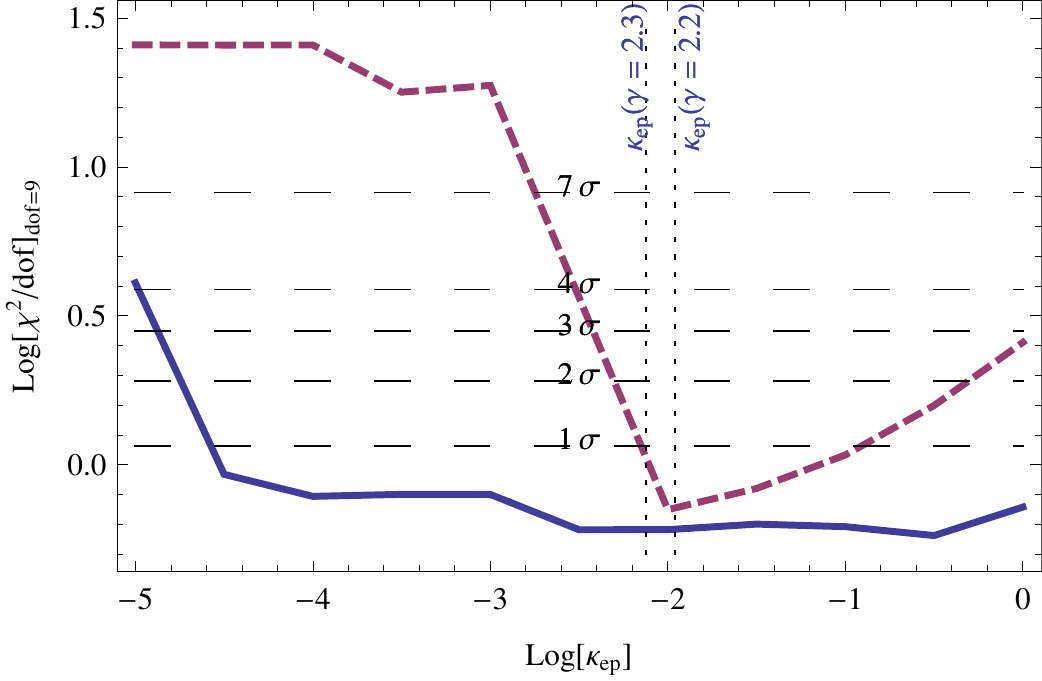,width=\columnwidth}
\caption {~$\chi^2/dof$ as a function of $\kappa_{ep}$, the at-injection ratio of electrons to protons at $E = 1$ TeV ($dof = 9$). 
The {\bf solid curve} is for a $\chi^2$ obtained from a pure fit to the broadband data as described in the text ($dof$ = 9).
The {\bf dashed   curve} requires the wind speed obey Eq.~\ref{eqn_ZV_vel} with $\eta_{\rm therm}$ a free parameter (constrained to lie in the range $0.1 < \eta_{\rm therm} < 0.9$)  with the constraint that the total non-thermal pressure (generated by the modelled magnetic field and CR population) be no larger than the pressure launching the wind as given by Eq.~\ref{eqn_Pc} ($dof$ = 9). 
The vertical dashed lines shows the ratio predicted \citep{Bell1978} when the injection spectra for both electrons and protons are pure power-laws-in-momentum
with spectral indices $\gamma = 2.2$ and 2.3 (consistent with the observed TeV spectral index).
}
\label{plotChiSqrdkappaep}
\end{figure}

Fig.\ref{plotChiSqrdkappaep} shows that fits to the broadband spectrum of the region only poorly constrain the at-injection ratio of electrons to protons (at 1 TeV)
so that, for instance, good fits are possible for either of the limiting cases where, on the one hand, equal {\it numbers} of CR electrons and protons are injected and, on the other hand, equal {\it power} is injected in CR electrons and protons.
As for other parameters, however, the imposition of additional, physically-motivated constraints (as described in the figure caption) 
restrict the parameter space somewhat.
It is certainly of note that the $\chi^2$ is minimized for $\kappa_{ep}$ in the range predicted \citep{Bell1978} 
for the case that the injection spectra for both electrons and protons are pure power-laws-in-momentum
(and equal {\it numbers} of protons and electrons at thermal energies are injected into the acceleration process)
with spectral indices in the range $\gamma = 2.2-2.3$  (consistent with the observed TeV spectral index); admittedly, however, 
the $2\sigma$-allowed parameter space of this parameter is still fairly large.
Also consonant with the usual picture \citep[e.g.,][]{Thompson2006}, Fig.~\ref{plotChiSqrdPower} shows that CR electrons take around
10\% of the total CR power for the best-fit region of parameter space.

\section{Discussion: Implications of Modelling Results}
\label{sctn_Implications}

\subsection{Cosmic ray heating and ionization of GC gas}

On the basis of our modelling, we find that the CRs in the HESS region induce a local ionization rate of $\gtrsim 10^{-15}$ s$^{-1}$.
This is precisely in the range required by the data discussed in Appendix \ref{section_GCISM}.

Moreover, on the basis of the results obtained by \citet{Suchkov1993} for star-bursting environments (see their Fig.~1), this ionization rate would keep the GC molecular gas at a temperature of $\gtrsim$50 K, again precisely in the range required by the data \citep[also see][]{Yusef-Zadeh2007,Dogiel2009_III}. 
On the other hand, given this temperature, the kinetic pressure in the GC region would then imply that the molecular material would typically be at $\sim 10^5$ cm$^{-3}$, a number density higher than that found (or allowed) by our fitting (cf. fig \ref{plotChiSqrdnH}), which is typically much more tenuous.

A compelling resolution of these facts is that the molecular gas exists, in fact, in two phases that are in pressure equilibrium at $\sim 5 \times 10^6$ K cm$^{-3}$: 
dense cores of $10^5$ cm$^{-3}$ and relatively tenuous envelope material.
Note that in this scenario pressure equilibrium is established between the kinetic pressure of the core material and the virial pressure implied by the turbulent motions of the tenuous gas. Also note that for a typical GC gas turbulent velocity dispersion of 20--40 km/s, pressure equilibrium implies the tenuous material has a density in the range $50 - 200$ cm$^{-3}$, comfortably within the range given by our modelling. Finally, both phases will be heated by CR ionization to similar kinetic temperatures {\it assuming} the low-energy heating/ionizing cosmic rays can actually penetrate into the cloud cores.
As we have already seen, however, such penetration is extremely unlikely.
This fact, together with the higher cooling efficiency of the dense core material \citep[see fig 1 of][]{Suchkov1993}, suggests that it should  have a lower
kinetic temperature than the envelope material.
 This is all precisely in line with the phenomenology discussed in Appendix \ref{section_GCISM}:
 the GC molecular gas does indeed seem to be in a two-temperature distribution \citep{Oka1998,Rodriguez-Fernandez2001,Goto2008,Oka2005,Yusef-Zadeh2007} with warm, low-density gas enveloping cooler, denser cores.

The exclusion of CRs from the densest molecular core regions suggests CRs do not significantly heat or ionize the ambient, dense gas where stars actually form.
This is an interesting conclusion to reach in the context that the GC be considered a  true star-burst analogue;
it has recently been argued that CR heating and ionization 
of molecular gas in star-burst environments
radically alters the conditions for star-formation 
(away from those encountered in more quiescent star-forming regions) thereby leading to a top-heavy IMF \citep[cf.][]{Papadopoulos2010,Papadopoulos2010b}.
This mechanism would seem to be excluded in the GC.
Whether it might still operate in `true' star-bursts \citep[which, e.g., are dense enough that calorimetry is putatively achieved:][]{Thompson2006} we cannot yet say. 
We do note, however, that turbulent velocities typical for the most luminous starbursts \citep[which may exceed 200 km/s:][]{Downes1998} may imply advection timescales {\it into} molecular clouds similar to $t_{\textrm\tiny{wind}}$ in these systems (cf. the GC case, \S\ref{sctn_Medium}), 
potentially implying CRs in these clouds do reach into the dense gas \citep[cf.][]{Contini2010}.

\subsection{GeV emission from the GC}

It is clear from the broad-band spectrum presented in Fig.~\ref{plotBrdBndSpctraBestFit} that we explain $\lesssim10$\% of the observed $\gamma$-ray emission at $\sim$GeV energies.
This is to be expected given, as remarked above, that we only have access to total flux data from Fermi and that these receive a substantial contribution from discrete sources in the GC \citep{Chernyakova2010} and from both discrete and diffuse emission sources along the line of sight. 

\subsection{Explaining DNS/GC lobe radio emission}

The observational study by \citet{Law2010} revealed that the $\sim$GHz spectral index of the GC lobe steepens with increasing distance (both north and south) of the Galactic plane. 
This constitutes strong evidence for synchrotron ageing of a CR electron population  transported out of the GC, entirely consistent with the wind scenario we have been postulating here. 
Also of note is that the vertical extension of the GC lobe is very similar to the DNS. 
We previously postulated (Paper II), therefore, that the DNS and GC lobe are essentially identical structures synchrotron-illuminated by CR electrons advected from the inner GC (essentially the HESS region) on a wind \citep[cf.][on the radio halo of M82]{Seaquist1991}.

Three basic conditions must be satisfied in order that electrons escaping from the HESS field go on to pervade the DNS region and generate most of its synchrotron emission at $\sim$GHz wavelengths, viz.: i) the spectrum of the electrons leaving the HESS region (i.e., the steady-state distribution in that region as given by Eq.\ref {solutionSmpl}) must match the spectrum {\it at injection} required for the DNS electrons;
ii) the power in electrons leaving the HESS region must be enough to support the DNS electron population; and
iii) the time to transport electrons over the extent of the DNS must be less than or of the order of the loss time over the same scale.
As shown in Paper II, all these conditions are well-met within our scenario.
Firstly, previous modelling by some of us \citep{Crocker2010} shows that electrons are injected into the DNS with a spectral index in the range 2.0--2.4, perfectly consistent with the spectrum of electrons {\it leaving} the HESS region.
Secondly,
the same modelling demonstrated the total power required to sustain the steady-state electron population of the DNS is $\sim$ few$ \times 10^{37}$ erg/s, again consistent with the power advected away by HESS region electrons.
Thirdly, we have also shown previously that the magnetic field of the DNS region has a (2 $\sigma$) lower limit at 50 $\mu$G and a probable value of $\sim$ 100 $\mu$G.
Requiring that electrons can be transported over the entire vertical extent of the DNS ($\sim 1^\circ$ from the plane, corresponding to 140 pc) implies a wind speed of a least a few hundred km/s (see Fig.~1 of Paper II), again consistent with the wind speed predicted by the modelling of the HESS region.
More detailed modelling of electron transport out of the central HESS region and the attendant synchrotron cooling in order to match the detailed, position-dependent radio spectrum of the lobe (cf. the studies by \citet{Zirakashvili2006} and \citet{Heesen2009} of NGC 253) we leave to future work.

\subsection{Explaining the WMAP `haze' and Fermi `bubbles'}
\label{sectn_hazes}

From detailed morphological and spectral analysis of large angular scale maps of microwave and $\sim$GeV $\gamma$-ray emission produced, respectively, by the WMAP and Fermi instruments, Finkbeiner and co-workers \citep{Finkbeiner2004_II, Dobler2008,Dobler2009,Su2010} have made the interesting claim that there is evidence for anomalous, non-thermal `hazes' of emission centred on the GC.
This emission is apparently also coincident with extended soft X-ray emission traced by ROSAT \citep{Snowden1997}.
The most recent analysis \citep{Su2010} shows the anomalous $\sim$GeV emission to resemble two bilateral ÔbubblesÕ of emission centered on the core of the Galaxy and extending to around ±10 kpc above and below the Galactic plane.

As has been explored at length elsewhere \citep{Crocker2010b}, the WMAP haze and the Fermi bubbles are naturally explained within our model as due to the population of relic cosmic rays protons and heavier ions (not CR electrons as has mostly been explored in the literature), injected in association with the high areal density star-formation in the Galactic center.
In this scenario, the {\it secondary} electrons created by the hadronic collisions of these ions -- occurring within the rarefied wind environment -- naturally supply the synchrotron emission putatively detected by WMAP. 
The accompanying neutral meson decay (with a sub-dominant contribution from the IC emission by the secondary electrons), on the other hand, naturally explains the Fermi emission.
This explanation requires that these cosmic ray hadrons i) carry off a power of at least $10^{39}$ erg/s and ii) have been injected
over an extremely long timescale, at least 5 Gyr (the latter in order that $pp$ energy losses in the extremely rarefied outflow, $n_H \sim 0.01$ cm$^{-3}$ put the local CR ion population into saturation).

An important result of this current work -- in support of the explanation of the large-angular-scale microwave and GeV emission we advocate -- is the demonstration that objects inhabiting or processes occurring within the HESS region do, indeed, accelerate $\gtrsim 10^{39}$ erg/s in cosmic rays (cf. Fig.~\ref{plotChiSqrdPower}).
We have made the case here that GC supernovae can supply the power required to sustain the acceleration of this non-thermal particle population but note that this determination (that $\gtrsim 10^{39}$ erg/s is injected into CRs) is independent of uncertainties surrounding the supernova rate, efficiency of SNRs as cosmic ray accelerators in the GC environment, etc.
Note, in fact, that we cannot exclude, in principle, the possibility that other sources in the region might ultimately supply the power required by the CRs; the super-massive black hole Sgr A$^*$ is an obvious candidate \citep{Aharonian2006} particularly as recent analysis \citep{Chernyakova2010} of the GeV and TeV point sources coincident with it suggest that it, too, roughly supplies of order $10^{39}$ erg/s in cosmic rays.

The long timescale of injection required to explain the haze/bubbles would -- were it associated with any other Galactic source of structure -- immediately call this explanation into question.
The morphology of the observed bubbles \citep{Su2010}, however, unambiguously associates them with the GC, probably the only structure in the Galaxy for which spatially-localized star formation over such a long period is credible \citep{Serabyn1996,Figer2004}.

Another interesting aspect of this scenario is that the wind of plasma and CRs leaving the GC should neither escape to infinity nor fall back to the Galactic plane.
In fact, given a wind speed of few hundred km/s,
it is probable that the wind injected flow out of the GC stalls at a height of a few kpc -- just as required -- as has recently been investigated at length by \citet[see Appendix \ref{section_Stalling}]{Rodriguez-Gonzalez2009}.

\subsection{Galactic bulge positrons}

A nuclear wind could also help disperse positrons into the Galactic bulge \citep[cf.][] {Dermer1997,Casse2004,Prantzos2006,Totani2006,Higdon2009,Gebauer2009,Churazov2010}, 
potentially 
explaining the morphology of the observed 511 keV emission line  \citep{Knodlseder2005,Weidenspointner2008} as we also explain in Paper I.
It should be noted, however, that this postulate does not address the problem of where the positrons come from in the first place.

A prime candidate process to supply the requisite, low-energy positrons is
 inverse beta decay of radioisotopes 
injected by GC Type 1a supernovae.
It has been claimed, however, that  GC SN1a can, only supply 10\% of the positrons required by observations \citep{Schanne2007}.
This determination assumes 3.3\% of the positrons from $^{56}$Ni decay escape the Type 1a envelope.
This fraction, however, obtained from normalizing to a statistical sample of optical light curves  \citep{Milne1999}, is model-dependent and depends, in particular, on assumptions about the magnetic field strength and topology and the state of the gas in the expanding remnant
 \citep{Milne1999,Higdon2009,Martin2010}.
Indeed, higher values for it have been countenanced \citep{Higdon2009} and direct upper limits are at the $\gtrsim$ 12\% level for a number of {\it disk} SNRs 
\citep{Martin2010}.
In any case, the highly porous and, in parts, extremely rarefied GC environment may allow many more positrons to escape the envelopes of GC SN Ia's  than typical for disk-type environments \citep[cf.][]{Parizot2005,Higdon2009,Martin2010}.
This consideration may extend to other classes of SN in the GC environment.

The many massive, young stars in the GC 
also constitute a compelling source population for an unusual sub-class of SN Type 1c, thought to originate in Wolf-Rayet explosions and exemplified by GRB-associated SN  2003dh, that have been speculated \citep{Casse2004} to be sources of the bulge positrons, as have GRBs more generally \citep{Bertone2006,Parizot2005}.
Another possibility is that processes unique to the GC may be acting to supply positrons \citep[e.g., tidal disruptions of massive stars by the SMBH:][]{Cheng2006,Totani2006} as, indeed, may the population of millisecond pulsars \citep{Wang2006} that the GeV spectrum of the region indicates may be present.
We leave investigation of these speculations to further work.

As a final note in this section, \citet{Churazov2010} have recently concluded (on the basis of a spectral study of SPI/INTEGRAL data on the circum-GC 511 keV line emission) that  the annihilating positrons are injected into a hot ($\sim 10^6$~K), volume-filling ISM plasma phase which undergoes radiative cooling.
The initial temperature of this gas is somewhat cooler than favoured by our analysis but we do note that, broadly, this scenario would seem to be consonant with our picture.
It is also of interest that the prior modelling by \citet{Rodriguez-Gonzalez2009} of a star-formation-driven GC outflow with initial temperature $\sim 10^7$ K shows that the plasma cools to $\sim 10^5$ K by the time it has reached a height of $\sim$ kpc.
According to the modelling of \citet{Churazov2010} at such a temperature the positrons may annihilate so, again, the combination of these two ideas would seem to point to a model that matches the observed $\sim$kpc extension of the 511 emission.

\section{Conclusions}
\label{sctn_Conclusions}

We have modelled in detail the high-energy astrophysics of the inner $\sim$200 pc of the Galaxy with the aim of explaining a number of observations including the region's non-thermal radio and $\sim$TeV $\gamma$-ray emission.
On the basis of our modelling/analysis we determine the following:-

\begin{enumerate}

\item Supernovae occur in the inner $1^\circ.6$ (in diameter) of the Galaxy at a rate of approximately 1 per 2500 years and release $\sim 1.3 \times 10^{40}$ erg/s in mechanical power (which is supplemented by massive stellar winds at the $\sim$10\% level; see Appendix \ref{sctn_SFR}).
This is precisely enough to drive the fast outflow ($\gtrsim 100$ km/s) from the region and support the region's non-thermal particle population.
Given
the data, moreover, are consistent with the current star-formation and supernova rates being close to their long-term average values, a system in quasi-steady-state is suggested.
%and, moreover, the {\it current} supernova rate is sufficient to sustain them. 
%
 
 \item Despite the GC resembling a star-burst in many ways, a number of the circumstances \citet{Thompson2006} claim for star-bursts apparently do not pertain in the GC:
 
\begin{enumerate} 
 \item We conform that, in general,  `calorimetry' fails in the GC: $\gtrsim 95$ \% of the power injected into non-thermal particles is advected from the system, viz.~$\sim 10^{39}$ erg/s. This despite the high surface mass density of gas and the relatively high magnetic field amplitude \citep[cf.][]{Thompson2006}. Also of note is the fact that particle transport is overwhelmingly advective; given, presumably, the strong, turbulent magnetic fields of the region, particle diffusion timescale are much longer than either advection or energy-loss timescales for both protons and electrons. This severely limits the role of the super-massive black hole on larger angular scales consistent with the determination that supernovae and stellar winds can drive the high-energy activity
(setting aside the $\gamma$-ray point source coincident with Sgr A$^*$).
 
\item Despite the high volumetric-average gas density, secondary electrons probably play only a minor role in generating the observed radio emission from the GC.
Both pure leptonic and pure hadronic scenarios for explaining the broadband emission of the HESS region are rejected with reasonable confidence, mirroring the conclusions of \citet{Crocker2007} on the Sagittarius B region, one of the largest giant molecular cloud complexes in the CMZ.

\item The large-scale magnetic field of the GC is bounded by 60 $\mu$G $\gtrsim$ B$ \gtrsim$ 400 $\mu$G. 
While certainly strong with respect to the typical Disk field amplitude and, in fact, comparable to the field strengths determined for the central regions of nearby starbursts \citep[e.g., NGC 253:][]{Heesen2009b}, this is short of the strength required, {\it by itself}, to establish either electron calorimetry or
hydrostatic equilibrium (though, as has been noted, hydrostatic equilibrium does not seem to pertain in any case given the existence of the wind -- driven by the combined pressure due to {\it all} ISM components).

\end{enumerate} 

In this context, one should keep in mind that the extent to which
Thompson et al.'s starburst model pertains in reality to genuine starbursts is contested.
The recent TeV $\gamma$-ray detection of NGC253, e.g., 
has been interpreted by the HESS collaboration, on the one hand, as indicating
 this system falls well short of being calorimetric \citep{Acero2009} but \citet{Lacki2010} reach a different conclusion (though the latter authors also grant that the system does fall short of being fully calorimetric).
We also note with interest a very recent paper by \citet{Lacki2010b} investigating the high-energy astrophysics of a number of starburst-like environments including the GC (in the context of evaluating the contribution of electron synchrotron to the diffuse, soft X-ray emission detected from these systems; the authors find a negligible contribution in the GC case, a conclusion with which we agree: cf. Fig.~\ref{plotBrdBndSpctraBestFit}).

\item The GC magnetic field is also significantly weaker than the $\sim$mG amplitude predicted by analysis of the NTFs \citep{Yusef-Zadeh1987,Morris1989}.
On the other hand, it is likely very close to energy equipartition/pressure equilibrium with the very hot plasma (and turbulent motions of the dense molecular gas).
The best-fit magnetic field also falls very close to the value required to establish the phenomenologically-expected \citep{Thompson2006} scaling $U_B \simeq 2 \ U_{ISRF}$.
These facts tend to suggest the field is generated locally in the GC ISM through, e.g., a turbulent dynamo.
%

%Nevertheless, the strong RC evidence for an overall poloidal field geometry (outside the molecular phase) remains.
%
%This morphological evidence (together with the previously-claimed $\sim$mG field amplitudes) has previously been invoked \citep{Chandran2000,Morris2007} as evidence that the GC field  is  generated globally via the gradual accumulation of vertical magnetic flux
%through the accretion of partially-ionized disk gas into the nucleus over the lifetime of the Galaxy.
%
%Were this mechanism to operate, our magnetic field upper limits would imply that the accumulated seed field was approximately an order of magnitude weaker than that originally proposed \citep[which was $\gtrsim 2 \times 10^{-7}$ G:][]{Chandran2000}.
%

\item The modelling tends to favour regions of parameter space where the wind is driven approximately equally by the pressures ascribable to thermal (plasma) and non-thermal (magnetic field and CRs) GC ISM phases \citep[cf.][on NGC 253]{Zirakashvili2006}.
The wind has an effective density of $n_H \sim$ 0.03 cm$^{-3}$, a characteristic speed near the mid plane satisfying 100--1200 km/s and most probably 400--800 km/s and advects about 0.02-0.03 $\msun$/year of gas from the GC (see Appendix \ref{sctn_TotalMass}).

\item  Given the total power implied by our modelling and the region's SN rate, the simplest -- though somewhat surprising --  global explanation of the GC's non-thermal emission is that the region's SNRs are the dominant source of cosmic rays {\it which they accelerate with at least typical efficiency}. This despite the much larger volumetric-average higher-density ambient medium which would be naively expected to significantly alter the evolution of the SNRs \citep{Spergel1992}. 
This is consistent with the picture that the GC supernovae are exploding into a highly porous medium \citep{Reynolds2008,Ksenofontov2010} excavated in part by 
the progenitor star's wind \citep{Volk1988,Biermann1993},
stellar winds in general, and/or by previous supenovae \citep[cf.][]{Westmoquette2009}.

On the other hand, it should be borne in mind that the evidence associating GC SNRs with acceleration of cosmic rays is only circumstantial 
and, in fact, any process proportional to the supernova rate and producing an average $\sim 10^{50}$ erg per SN in non-thermal particles satisfies the phenomenology \citep[cf. ][]{Voelk1989b, Voelk1990}. Thus, it remains possible that at least some acceleration takes place, e.g., on general GC ISM turbulence injected by the collective activities of supernovae and stellar winds \citep{Drury2001}. 

It is also interesting that -- given the play afforded by observational uncertainty -- we have no strong evidence from the overall energetics and the FIR luminosity for (or against)  any strong biasing of the IMF of the GC stellar population \citep[cf.][]{Persic2010,Papadopoulos2010,Papadopoulos2010b}.

\item CR electrons accelerated in the HESS region lose $\lesssim$ 10 \% of their power to synchrotron emission there, generating thereby the diffuse, non-thermal radio emission detected across the same region.

\item Protons lose $\lesssim$ 1 \% of their power to $pp$ collisions on ambient gas in the inner few degrees.
These collisions explain the diffuse, TeV emission detected across the Central Molecular Zone (electrons only making a minor contribution to this emission). 
Interestingly, these $pp$ interactions occur in a relatively low density gas phase ($\lesssim 30 $ cm$^{-3}$) despite the fact that most ambient gas finds itself at much high density ($\gtrsim 10^4$ cm$^{-3}$) in the cores of giant molecular clouds. 
This is consistent with the picture that diffusive transport timescales in the GC (in this case, into the clouds) are much longer than advective ones.

\item The Coulomb collisions of the low energy members of the in situ, cosmic ray population (both protons and electrons) can inject the heat to maintain the temperature of the anomalously hot, relatively low-density molecular gas and supply much of the unusually high ionization power required by various observations.
On the other hand the molecular cores are at a lower kinetic temperature probably as a combination of their higher cooling efficiency and, presumably, the fact that even high energy CRs do not penetrate them during the time they remain in the region.
This seems to rule out a role for CR-modification to the region's dense gas chemistry \citep[cf.][]{Papadopoulos2010,Papadopoulos2010b}.

\item 
With plausible physical constraints, our modelling favours (at 2$\sigma$ level) CR proton energy densities in the approximate range $2.5$ eV cm$^{-3} < U_p < 50$ eV cm$^{-3}$ with a best-fit value around 20 eV  cm$^{-3}$. 
The allowed range of this parameter reflects the {\it a priori} uncertainty in the typical $n_H$ value where CR ion collisions are taking place.
The lower end of the acceptable $U_p$ parameter range -- corresponding to the upper end of the allowed gas density (i.e., $n_H \to \langle n_H\rangle_\textrm{\tiny{vol}}$) --
is only a factor of a few larger than the local CR energy density, and matches that found in the original analysis of the HESS group \citep{Aharonian2006}
where it was assumed that all of the molecular gas was `accessible' to the TeV CRs.

Likewise, our modelling favours (at 2$\sigma$ level) CR electron energy densities   in the approximate range $0.2$ eV cm$^{-3} < U_e < 4$ eV cm$^{-3}$.
The range of this parameter reflects the {\it a priori} uncertainty in the magnetic field amplitude where the synchrotron emission is occurring.

The existence of the fast wind is what ensures that the in situ CR density at the GC is probably only a few times larger than the solar value at $E_{CR} \simeq $ GeV -- despite an areal density of SNRs (and massive, windy stars) in the region much larger than typical for the Galaxy at large -- and also explains why the (non-thermal) radio continuum emission from the GC is smaller than expected on the basis of the region's total FIR luminosity and the FRC (Paper II).

\item %
We find that `equipartition' \citep{Beck2005} between the magnetic field and the non-thermal particle population 
can be satisfied over at least some of the parameter space favoured by our modelling.
Formally, good fits consistent with equipartition are possible up to above $10^{-4}$ Gauss or magnetic field and CR proton energy densities  surpassing 250 eV cm$^{-3}$.
Such fits, require, however very small ambient gas densities (1 cm$^{-3}$) and vanishing wind speed: these seem quite implausible to us in the light of all other evidence.
We think it likely that, in fact, equipartition does not hold in the GC ISM.

\item The relativistic electrons that escape the region  on the wind can explain the large-scale, diffuse radio emission detected from the Galaxy's nuclear bulge ($|l| < 3^\circ$), in particular their synchrotron emission explains the non-thermal spectrum of what has been labelled in previous work the DNS \citep{LaRosa2005} and the GC lobe \citep{Sofue1984,Law2010}.

\item We have not explained the majority of the apparently-extended $\sim$ GeV emission from the region but 
much of this emission is due to point sources certainly including a source coincident with Sgr A$^*$ \citep{Chernyakova2010}
with
preliminary indications that, not unexpectedly, millisecond pulsars also make a large contribution to the extended emission.

\item If -- as consistent with the work of others based on the current stellar population \citep{Serabyn1996,Figer2004,Figer2008} or upper limits to the pulsar population \citep{Lazio2008} -- the current SFR has been sustained for multi-Gyr timescales, the 
wind emerging from the
GC and the CR ion population it carries compellingly explain the large-angular-scale, non-thermal emission detected at microwave and $\sim$GeV energies.

\end{enumerate}

In summary, the following  picture emerges from the studies described here and in the preceding Papers I and II: the gradual accretion of gas into the Galactic nucleus -- presumably under the action of the Galactic stellar bar, possibly assisted by dynamical friction and magnetic viscosity effects \citep{Binney1991,Blitz1993,Serabyn1996,Morris1996,Beck1999,Stark2004} -- over timescales approaching the age of the Galaxy -- has led to sustained star-formation inside a region circumscribed by the Galaxy's inner Lindblad resonance.
Inside this radius lie the vast (but dynamic) gas reservoir represented by the CMZ and the nuclear stellar population resulting from the persistent star-formation.
Concomitant with this star-formation activity are supernovae, whose remnants 
accelerate the GC cosmic ray population and heat the region's hot plasma; 
the tell-tale, diffuse TeV $\gamma$-rays and X-rays produced by these two 
extend across the same inner region of the Galaxy.

Another by-product of the star-formation is the large-scale super-wind that emerges from the GC.
This acts to keep the energy density of the various GC ISM components in check, in particular, 
advecting CRs to very large scales above and below the Galactic plane.
Acting to also advect the magnetic field, the wind may well also have a crucial role in forming the NTFs \citep{Shore1999,Boldyrev2006} which, intriguingly, are also only detected over the same inner region of the Galaxy as the phenomena referred to above.
In the picture of \citep{Shore1999}, the NTFs
do not trace a global poloidal field topology,
but instead  are {\it dynamic} structures, akin to cometary plasma tails, formed from the interaction of the large-scale, magnetized plasma outflow draping the region's dense molecular clouds. 
Consonant with the above,  the latest NIR polarimetry measurements  \citep{Nishiyama2010} apparently reveal a global field topology that is toroidal in the plane (for $| l | < 2.0^\circ, | b | < 0.4^\circ$, closely corresponding the HESS region), bending to almost perpendicular for $| b | > 0.4^\circ$, presumably in the wind zone. 

Finally, 
although it should be remembered that, of course, some stochastic variation in the background star-formation rate is expected, we have pointed out here that the {\it current} star-formation rate in the GC seems to be close to its long-term average value.
This apparent stability suggests the operation of self-regulation or feedback mechanisms in which many of the actors discussed here -- the magnetic field, the CR population, the wind -- undoubtedly play some part, but we leave the detailed working-out of this to further work.

\section{Acknowledgements}

The authors gratefully acknowledge useful conversation or correspondence with Csaba Bal\'azs, Rainer Beck, Nicole Bell, Joss Bland-Hawthorn, Valenti Bosch-Ramon, Michael Brown, Sabrina Casanova, Masha Chernyakova, Roger Clay, John Dickey, Ron Ekers, Katia Ferri{\` e}re, Stanislav Kel'ner,  Mitya Khangulyan, Jasmina Lazendic-Galloway, Mark Morris,   Giovanni Natale, Emma de O\~na Wilhelmi, Ray Protheroe, Brian Reville, Frank Rieger, Ary Rodr{\'{\i}}guez-Gonz{\'a}lez, Todor Stanev,  Andrew Taylor, and Mark Wardle. DIJ thanks Gavin Rowell for providing the original version of the script with which the Fermi raw count map was made. RMC particularly thanks  Heinz V{\"o}lk for many enlightening discussions.

%\begin{thebibliography}{99}

\clearpage

\appendix

\section{The GC ISM} 
\label{sectn_Inputs}

\subsection{Gas distribution}
\label{section_GCISM}

The interstellar medium throughout the GC region consists of a number of co-existing phases with very different physical parameters
which we now partially describe.
For full reviews -- particularly of the H{\sc i} and relatively cooler plasma phases not described below -- the reader is referred to the work of
 \citet{Ferriere2007} and \citet{Higdon2009}. 

\subsubsection{Molecular gas}

In its molecular phase, the inner $\sim$ 500 pc of the Galaxy contains $\sim$5\% of the Galaxy's allocation of molecular hydrogen \citep[a total mass of a few times $10^7 \, \msun$:][]{Ferriere2007}, mostly within the asymmetric condensation along the Galactic plane known as the Central Molecular Zone
 \citep[CMZ;][]{Morris1996}. 
The overall geometry of the CMZ structure -- seen in projection -- is very difficult to reconstruct. 
Much of the dense molecular gas seems to lie on a $\sim$180 pc radius ring around the GC with a relatively tenuous interior  \citep[see][and references therein]{Morris1996,Launhardt2002,Liszt2009}, akin to the nuclear star-forming rings commonly observed in barred spiral galaxies nominally similar to our own.
Inside this radius a number of giant molecular cloud complexes are to be found, including Sgr B2, the most massive in the Galaxy \citep[see][and references therein]{Crocker2007}.

Aside from these morphological issues, the CMZ is characterised by:  
\begin{itemize}

\item Unusually high inter- and intra-molecular clump velocity dispersions of 15-30 km/s \citep{Oka1998,Gusten2004}, ranging to even higher values $> 50$ km/s in particularly active regions like Sgr B2 \citep[see, e.g.,][and references therein]{Crocker2007}; cf. typical disk cloud velocity dispersions of $\sim$ 5 km/s.

\item Unusually high molecular densities: a typical H$_2$ molecule will find itself in an ambient density of $\sim 10^4$ cm$^{-3}$ \citep{Paglione1998}; cf. typical disk cloud densities of $\sim 10^2$ cm$^{-3}$ -- though it should be emphasised that the filling factors of the extremely dense molecular cloud cores are extremely small, $\lesssim 0.01$ \citep{Oka2005}, and the gas distribution is likely highly fragmented \citep[rather akin to the situation within luminous star-bursts: e.g.,][]{Westmoquette2009}.

\item Unusually high kinetic temperatures (molecular material is typically found with temperatures in the range 50-70 K but co-existing phases with temperatures of up to 300 K are detected; cf. typical disk cloud temperatures of $\sim$ 15 K) and virial pressures (of $\sim 5 \times 10^6$ K cm$^{-3}$ -- cf. that typical for the Galactic disk, $\sim 2 \times 10^4 $ K cm$^{-3}$; \citealt{Spergel1992}). 

\item Pervasive shocks, as revealed by  thermal SiO emission \citep{Martin-Pintado1997,Huettemeister1998} and -- as revealed by surveys of molecular lines and dust continuum emission at mm and sub-mm wavelengths \citep{Oka1998,Tsuboi1998,Pierce-Price2000,Oka2010} -- filaments, arcs, and shells, indicative of local, turbulent sources \citep{Tanaka2007} and explosive events.
\end{itemize}

In addition to the dense molecular cores, in the last few years evidence for another, comparatively diffuse ($\sim 100$ cm$^{-3}$) and hot ($\sim$ 250 K) molecular phase has emerged from 
comparison of the intensity of different emission lines of $^{12}$CO \citep{Oka1998}, other molecules  \citep{Rodriguez-Fernandez2001},  and from
spectroscopic observations of  line absorption by $H_3^+$ \citep{Goto2008,Yusef-Zadeh2007,Oka2005} at infrared wavelengths. 
This environment,
which represents $\sim$30\% of the total molecular gas by mass \citep{Ferriere2007}, appears to be unique within the Galaxy to the GC and requires a distributed heating mechanism. 
The large $H_3^+$ columns observed towards the GC furthermore imply an ionization rate of $\gtrsim 10^{-15}$ s$^{-1}$, much higher than typical in the diffuse ISM of the Galactic disk ($\sim 3 \times 10^{-17}$ s$^{-1}$).
These same observations, moreover,  show -- contrary to the expectation from chemical model calculations -- that the relative number density of  $H_3^+$ is much higher in diffuse clouds than in dense clouds towards the GC.
In contrast to the gas temperature, dust temperatures throughout the CMZ are anomalously low at around 20 K \citep{Pierce-Price2000}, suggesting that a mechanism for gas-heating is required for the CMZ unrelated to ultra-violet radiation.

\subsubsection{Hot plasma and X-ray observations}
\label{sctn_Plasma}

In its plasma phase, X-ray continuum and Fe line observations apparently reveal a two-temperature plasma containing `hot' ($\sim$ 1 keV) and (mysteriously) `very hot' (6-9 keV) components \citep{Koyama1989,Yamauchi1990,Kaneda1997,Muno2004,Belanger2004}. 
The X-ray emission from the putative very hot component is strongly concentrated within the inner $\sim$150 pc (in diameter) of the Galaxy \citep{Yamauchi1990,Belmont2005}. 
As first observed by \citet{Spergel1992}, 
there may be pressure equilibrium \citep[at $(3-6) \times 10^6$ K cm$^{-3}$:][]{Koyama1996,Muno2004} between the kinetic pressure of the putative
very hot plasma phase  and the virial pressure implied by
the  
turbulent motions of the molecular gas.
%Some researchers have, however, have long argued that the apparently diffuse emission is actually due to unresolved, faint X-ray sources. 
%Recent observations with Chandra, however,  have shown\citep{Revnivtsev2009} that the X-ray emission 
The very-hot plasma presents a severe energetics problem, however: were it a hydrogen plasma its sound speed would be $\sim$1500 km/s -- considerably in excess of the local escape velocity of $\sim$ 900 km/s \citep{Muno2004} -- and it would escape \citep{Yamauchi1990} on a short timescale. 
A steady-state situation would require a power to sustain this outflow considerably in excess of $10^{40}$ erg/s.
A second difficulty is that  there is no widely-accepted mechanism to heat the plasma to more than a few keV.

There is no universally-accepted resolution to these anomalies. One interesting suggestion is that the 8 keV emission is due to a very hot {\it helium} plasma which would be gravitationally bound \citep{Belmont2005}.
Another suggestion is that the `plasma' is illusory, the emission actually being attributable to unresolved point sources.
Recent Chandra observations around $l$ = 0.08, $b$ =1.42  (taken to be typical of the so-called X-ray Ridge) support this sort of picture \citep{Revnivtsev2009}.
On the other hand, the situation within the inner $\sim$150 pc -- where the 6.7 keV Fe line emission strongly peaks \citep{Yamauchi1990} --
may be quite different to that pertaining elsewhere in the Galaxy \citep{Dogiel2009_II}.
A deep observation \citep{Muno2004} of the inner 17$'$ with Chandra could explain only $\lesssim 40 \%$ of the X-ray flux as due to dim point sources.
Moreover, recent results obtained with the SUZAKU X-ray telescope continue to clearly suggest \citep{Koyama2007,Dogiel2010} the existence of a hot plasma covering at least the central 20$'$; this issue, therefore, has remained unresolved.
%
%Below we put forward a number of arguments in support of the existence of a very hot plasma

\subsection{The GC magnetic field}
\label{sctn_GCBFld}

The amplitude of the {\it large-scale} magnetic field threading the GC region has remained uncertain for about twenty-five years \citep[see][for a recent review]{Ferriere2009}. 
Three different scales of field amplitude have been suggested in the light of different analyses:
\begin{itemize}

\item {\bf mG field:} $\sim$GHz radio observations of the region reveal the presence of so-called ``non-thermal filaments" \citep[NTFs; see ][for a review of these fascinating structures]{Bicknell2001}, long and thin magnetic flux tubes that run predominantly perpendicular to the Galactic plane through the central $\sim 200$ pc (diameter) and which are illuminated by synchrotron emission from relativistic electrons. The rigidity of the NTFs requires that they are suffused by magnetic fields of $\sim$mG amplitude \citep{Yusef-Zadeh1987}. 
A countervailing external pressure is then required lest the NTFs dissipate over short timescales given their large, internal magnetic pressures and, given observational constraints, this can only be supplied by an external magnetic field also of $\sim$ mG strength. 
This implies a $\sim$ mG field pervades the region \citep{Morris1989}.

\item {\bf 100 $\mu$G field:} Alternatively, the assumption of
pressure equilibrium between the various phases of the Galactic Centre
interstellar medium (including turbulent molecular gas; the
contested \citep{Revnivtsev2009} ``very hot" plasma;
and the magnetic field) suggests fields of $\sim$100 $\mu$G over $\sim$400
pc (diameter) size scales \citep{Spergel1992}.
%
%Studies at mm wavelengths \citep{Fukui2006} also suggest that the field within $\sim$ 1 kpc of the GC is $\sim$ 100 $\mu$G. 

\item {\bf 10 $\mu$G field:} Given their detection of the DNS source alluded to above, \citep{LaRosa2005} have invoked the ``equipartition" condition \citep[minimizing the total energy in magnetic fields + cosmic rays, incidentally predicting rough equipartition in energy density between these two:][]{Burbidge1956,Beck2005} to arrive at a large-scale field (over the central $\sim$800 pc in diameter) amplitude of $\sim$10
$\mu$G within which cosmic ray electrons are synchrotron-radiating to produce the observed, non-thermal radio emission. 
\end{itemize}

As remarked above, some of us \citep{Crocker2010} were recently able to show -- with reference to the higher-frequency extension of the DNS radio spectrum in concert with GeV $\gamma$-ray data -- that magnetic field amplitudes lower than 50 $\mu$G are excluded over the central $\sim 800$ pc (in diameter). 
This work also calls into question the {\it assumption} of equipartition which may fail in the GC as well as in other star-forming regions \citep{Thompson2006}. 
Other pieces of evidence -- as described below -- support the notion that the real DNS field is $\sim$ 100 $\mu$G. (This still implies an energy density more than 100 times larger than that typical for the field in the Galactic disk.) 

\subsubsection{Magnetic field required for equipartition with light field}
A magnetic field of $\sim 60 \ \mu$G would be required were equipartition with
the volumetric-average $U_\textrm{\tiny{ISRF}} \simeq 90$ eV cm$^{-3}$ over the HESS region to be satisfied. 
In the Galactic plane -- and many other astrophysical environments \citep[see, e.g.,][]{Thompson2006}-- it can be inferred that the magnetic field and interstellar radiation field globally satisfy $U_B \sim 2 \ U_\textrm{\tiny{ISRF}}$. 
A close-to-constant ratio between the two quantities certainly seems to be necessary to maintain the linearity of the FRC \citep[][]{Lisenfeld1996}. 
There are dynamic, local mechanisms that have been put forward to explain why this scaling should apply. These rely on the notion that magnetic fields arise in turbulent dynamo action with the ultimate source of 
such turbulence being the local, massive stellar population
which injects turbulence via stellar winds and SN explosions \citep{Lisenfeld1996} or, more directly, through radiation pressure \citep[in starburst environments:][]{Thompson2008}.

\subsubsection{Magnetic field required for hydrostatic equilibrium}
\label{section_Bfield}

A magnetic field exceeding that required to establish hydrostatic equilibrium, $B_{eqm}$, should lead to the development of the \citet{Parker1966} instability and thus  $B_{eqm}$ might be considered an upper limit within the scenario explored by  \citet{Thompson2006} and \citet{Lacki2009}.
The field required for hydrostatic equilibrium can be estimated as $2 \sqrt{\Sigma_{gas} \Sigma_{tot}}$ mG where the gas and total surface mass densities are given in g cm$^{-2}$ \citep{Thompson2006}.
The average surface mass density in {\it gas} over the HESS and DNS regions can be estimated on the basis of the work of \citet{Ferriere2007} to be 0.07 g cm$^{-2}$ and 0.01 g cm$^{-2}$ respectively. The {\it total} surface mass density -- dominated by stars -- we infer from the work of  \citet{Launhardt2002} to be around 21  times the gas surface density for the HESS region and 36 times for the DNS region. 
Hydrostatic equilibrium therefore requires fields of $\sim$ 600 $\mu$G and $\sim$ 200 $\mu$G for the HESS and DNS regions.
It is noteworthy, then, that our previous work points to a field approaching this strength in the DNS.
Interestingly, a Parker-like instability is suggested by mm-wave observations \citep{Fukui2006} of molecular filaments of several hundred parsec length within $\sim$ 1 kpc of the GC (these measurements also suggest a $\sim$ 100 $\mu$G field amplitude). 
These observations have independently been interpreted as suggesting a $\sim$ 100 $\mu$G field \citep{Fukui2006,Crocker2010}.

\subsubsection{Magnetic field in GC molecular clouds}

Measurements of the topology and amplitude of the magnetic field interior to the GC's giant molecular cloud population have been obtained via  
sub-millimetre polarimetry \citep{Novak2003,Chuss2005} and Zeeman mapping of the H I line in absorption \citep{Crutcher1996}. 
These measurements show that the cloud-interior fields are drawn out along a direction following the ridge of molecular and thermal radio emission, predominantly parallel with the Galactic plane. Such a configuration might arise naturally from the shearing action of the tidal field at the GC on the orbital motion on the molecular gas.
Field amplitudes, moreover, reach $\sim$mG  levels within, e.g., the massive molecular envelope surrounding the dense Sgr B2 star-forming cores \citep{Lis1989,Crutcher1996} 
with lower limits on larger scales at the $\sim$150 $\mu$G level \citep{Novak1997,Chuss2005}.

\subsubsection{Magnetic field topology}

The determination of the GC magnetic field's global topology is an area of active research.
The geometry of the NTFs implies not only that the field is strong, but also that it is regular and poloidal \citep[and references therein]{Morris2007,Ferriere2009}.
On the other hand, the field in dense molecular material as revealed by the techniques mentioned above is predominantly parallel to the plane,
at right angles to the large-scale  poloidal field structure {\it apparently} traced by the NTFs.
Most recently, NIR polarimetry measurements \citep{Nishiyama2010} -- in principle, sensitive to magnetic field structure in both dense and rarefied ISM phases -- have apparently revealed a global magnetic field structure that is parallel to the plane for Galactic latitudes $| b | < 0.4^\circ$ but bends to run perpendicular at higher latitudes.

\subsection{Volume of the source regions}

For the purposes of our modelling, motivated by the TeV and RC observations, for the
HESS region we assume a cylindrical volume  (seen in projection) which implies a volume of $9.7 \times 10^{61}$ cm$^{-3}$.
We assume the diffuse non-thermal source (DNS) emission volume to be an elliptical spheroid with circular cross-section in the Galactic plane (i.e., its greatest line-of-sight extent is equal to its width along the plane). This implies a total volume of $\sim 3.0 \times 10^{63}$ cm$^3$. 

\section{Star Formation and Supernovae in the GC}
\label{sctn_SFRplusSN}

For an initial estimate of the star-formation rate (SFR) in the GC, we turn to the work of \citet{Figer2004}. 
From this we infer that the HESS field is responsible for $\sim$ 2\% of the Galaxy's massive star formation \citep[nominates the region inside $r = 500$ as responsible for 10\% of the Galaxy's star formation]{Figer2008}. Given the SFR of the entire Galaxy is $\sim 4 \msun$/year \citep{Diehl2006}, the HESS field SFR is $\sim 0.08 \msun$/year.

\subsection{GC star formation rate}
\label{sctn_SFR}

\subsubsection{Star-formation rate from FIR emission}
\label{sctn_SFR_FIR}

A more accurate handle on the SFR is potentially given by its FIR emission. On the basis of the work of \citet{Kennicutt1998} we have
\be
SFR = \frac{L_{TIR}}{2.2 \times 10^{43} \ \textrm{erg/s}} \msun/\textrm{yr}
\ee
where the  total infra-red (TIR) emission is given by $L_{TIR} \simeq 1.75 \  L_{FIR}$; 
$L_{FIR} = L_{60 \ \tiny{\mu \textrm{m}}}[1 + S_{100 \ \tiny{\mu \textrm{m}}}/(2.58 \ S_{60 \ \tiny{\mu \textrm{m}}})]$; and 
$L(d= 8 \textrm{kpc})_{60 \ \tiny{\mu \textrm{m}}} \simeq 2.5 \times 10^{41}$ erg/s $S_{60 \ \tiny{\mu \textrm{m}}}/(10^6$ Jansky).

On the basis of IRAS data \citep{Launhardt2002}
the $L_{TIR}$ of the DNS source region is $2.8 \times 10^{42}$ erg/s, and that of the HESS field $1.6 \times 10^{42}$ erg/s[\footnote{We note  that the observed value of $L_{TIR} = 4.2 \times 10^8 \lsun$ for the HESS region is very close to the expectation generated by the total amount on $H_2$ in the system \citep[$8.7 \times 10^6 \msun$ as inferred from the data of][]{Ferriere2007} and the rough empirical correlation  between a given amount of (strictly, dense) $H_2$ and the IR emission of a system established by \citet{Mangum2008}:
\be
\log\left( \frac{L_{IR}}{\lsun} \right) \simeq 2.0 \ + \ 1.0 \ \log \left( \frac{M_{H_2}^\textrm{\tiny{dense}}}{\msun} \right)
\ee
on the basis of $H_2 CO$ observations of starburst galaxies.}], implying SFRs of 0.12 and 0.08  $\msun$/yr for the DNS and HESS regions, respectively, and consistent with the upper-limit on the SFR within the inner 400 pc quoted by \citet{Yusef-Zadeh2009} of 0.15 $\msun$/year (obtained by taking the stellar content of the region and an assumed age of $10^{10}$ years; also see \citet{Schanne2007}).
\citep[Note that young, massive Main-Sequence stars are the predominant contributors to the luminosity of both regions: see][]{Launhardt2002}.

%\subsubsection{Star-formation rate from radio continuum emission}

%implied $\Delta M$

\subsubsection{Interstellar radiation field}

We obtain the interstellar radiation fields (ISRFs) required by our modelling in the following fashion: i) for calculations concerning the DNS region, we employ the full interstellar radiation field determined for the Galaxy's inner 500 pc \citep{Porter2006}. This has an energy density of $\sim$ 19 eV cm$^{-3}$.
It is to be noted that -- as a check on self-consistency -- taking the volumetric average of the radiation energy density due to a point source of luminosity equal to $L_{TIR}^{DNS}$ we find 17 eV cm$^{-3}$, tolerably close to the result obtained from the work of \citet{Porter2006} (obtained for a somewhat different volume).

For ISRF of the HESS region, again using the $L_{TIR}$ of the region given above, we find a volumetric-average $U_\textrm{ISRF} \simeq 90$ eV cm$^{-3}$ over the HESS field; 
% and  $U_\textrm{ISRF} \simeq 19$ eV cm$^{-3}$ over the DNS region.
we assume the {\it shape} of the HESS ISRF spectrum is the same as that determined by \citet{Porter2006} for the inner 500 pc of the Galaxy.

\subsection{Supernova rate}
\label{section_SNrate}

A first approximation to the supernova rate is given by the consideration that the HESS field is responsible for $\sim$ 2\% of the Galaxy's massive star formation and that the Galaxy experiences $\sim$2 SN per century \citep{Diehl2006} so that we expect around 0.04 SN per century in the HESS field.
This simple estimate neglects effects like a possibly biased IMF in the GC; the
detailed and independent arguments we set out below, however, confirm a central value for the SN rate of 0.04 per century with an uncertainty of a factor of $\sim$2.
It is significant for our purposes (as we explain below) that the estimate we derive for the SFR from the FIR output of the region (which traces the {\it current} SFR) is consistent with other measures.

\subsubsection{Supernova rate from FIR emission}

We infer on the basis of the work of \citet{Thompson2007} that -- as both the supernova rate and the total IR (TIR) luminosity are proportional to the star formation rate -- one may relate the supernova rate to the $L_{TIR}$ via
\be
\frac{R_{SN}}{\textrm{century}} = 8.9 \times 10^{-11}  \ \beta_{17} \frac{L_{TIR}}{\lsun}
\ee
where $\beta_{17} \sim 1$, is described by \citet{Thompson2007}. With $L_{TIR} \simeq 4.2 \times 10^8 \lsun$ for the HESS field, we again determine a
supernova rate of 0.04/century.

\subsubsection{Supernova rate from stellar composition}

The type 1a supernova rate in the GC region defined by $r < 250$ pc and height $< 50$ pc has recently been estimated by \citet{Schanne2007} employing the scaling relations that have been determined \citep{Scannapieco2005} to relate the supernova rate to the star formation rate, $\dot{M}$ and total stellar mass of external galaxies, $M$:
\begin{equation}
\frac{R_{SN1a}(t)}{\textrm{century}} = A \ \frac{M_*(t)}{10^{10} \msun} \ + \ B \ \frac{\dot{M}_*(t)}{10^{10} \msun \textrm{Gyr}^{-1}}
\label{eqn_rateSN1a}
\end{equation}
where $A = 0.044 \pm0.015$ and $B = 2.6 \pm 1.1$ \citep{Schanne2007}. Adopting now the stellar mass of the Galactic nuclear bulge ($r < 300$ pc) determined by \citet{Launhardt2002} of $1.4 \times 10^9 \msun$, we can calculate the (central) value for the expectation of the SN1a rate in this region of the Galaxy to be 0.04/century  \citep[where we also assume a formation time of 10 Gyr;][]{Figer2004}. For an estimate of the core-collapse SN rate, we incorporate the fact \citep{Mannucci2005} that  the SN Ia rate is $0.35 \pm 0.08$ of the core-collapse rate in young stellar populations. Setting the first term in Eq.~\ref{eqn_rateSN1a} to zero, we determine that the expected core-collapse supernova rate in the nuclear bulge is 0.11/century (again taking central values) implying an overall rate of 0.15/century for all supernovae in the $r < 250$ pc region.

This result must now be scaled to the region under consideration (somewhat smaller than the nuclear bulge as defined above). 
On the basis of the data presented in table 7 of \citep{Launhardt2002}, there are two ways to do this: i) we can scale the above rate by the ratio of the inferred bolometric luminosity of the HESS field to that of the larger $r < 250$ pc region: $4 \times 10^8 \lsun/2.5 \times 10^9 \lsun = 0.2$; alternatively we can scale by the ratios of the inferred stellar masses of the two regions: $8 \times 10^8 \msun/1.4 \times 10^9 \msun = 0.6$. This implies a total supernova rate in the HESS field of 0.02--0.08/century.

Similar considerations to the above allow us to calculate that the SN rate within the entire DNS region (a volume $\sim$30 times larger than the HESS region) is  only $1.5-2.0$ times larger than in the HESS region alone. The SN activity is thus highly centrally peaked and, for simplicity, in our analysis we ignore the energy and particles injected by SNRs occurring in the DNS region but outside the HESS region.

\subsubsection{SN rate upper limit from GC pulsar population}

The rates found above are consistent with  that found by \citet{Lazio2008} and \citet{Deneva2009}
from their radio studies of pulsars and radio point source pulsar candidates in the inner 1$^\circ$ of the Galaxy, 
viz. that the GC SN rate has a rough upper bound at 0.1 per century (for a continuous -- rather than bursting -- star formation history, as we favour: see below).

\subsubsection{SN rate lower limit from X-ray emission}

We can arrive at a lower limit to the SN rate by requiring that there be enough power to support (only) the radiative loses of the plasma(s) and if we adopt the figure of 1\% of SN kinetic energy lost into plasma heating from \citet{Kaneda1997}.
With the total X-ray luminosity over the central region $\sim 4 \times 10^{37}$ erg/s \citep{Belmont2006}, this consideration defines a lower limit of 0.01/century.

\subsubsection{Gas turbulence}

The turbulent motions of the molecular gas throughout the region also represent a reservoir of energy which can be estimated to dissipate at a rate \citep{Bradford2005} $L_{turb} \simeq 0.42 v_{turb}^3/\Lambda_d$ where $\Lambda_d$ is a typical size scale for the turbulent structures. 
Estimating $v_{turb} \sim 25$ km/s and $\Lambda_d \sim 10$ pc we find $L_{turb} \sim 4 \times 10^{39}$ erg/s over the $\sim 9 \times 10^6$ solar masses of $H_2$ found in the HESS region. 

\subsubsection{High-velocity compact clouds}
\label{section_HVCCs}

A large-scale CO J=3-2 survey of the CMZ  with the Atacama Submillimeter-wave Telescope Experiment \citep{Oka2010}
detects an unusual population of high-velocity ($\Delta v \geq 50$ km/s) compact clouds (HVCCs) in the GC.
The kinetic power of  the HVCCs launched by explosive events in the HESS field is $3.8 \times 10^{39}$ erg/s, where the expansion velocity of each HVCC supplies the appropriate timescale to render a power from that HVCC's kinetic energy.

\subsubsection{Combined constraint; thermalization efficiency}
\label{section_Thermalization}

Considering the power required to sustain both gas turbulence and the HVCCs in the region, we can place a lower limit on the supernova rate in the region of $\sim 0.02$ per century assuming $\sim 100$ \% of the energy released per supernova goes into these two.
Turning these considerations around, again with the requirement that both turbulence and HVCCs be sustained and adopting the maximum supernova rate found above of 0.08/century we arrive at a conservative estimate of the minimum {\it overall} thermalization efficiency for GC supernovae of  $\eta_\textrm{\tiny{therm}}^\textrm{\tiny{min}} \equiv 0.4$
(where, in this instance, thermalization refers to all power lost into heating or moving the GC ISM rather than radiatively\footnote{Below, for purposes of determining the maximum and minimum star-formation-driven super-wind speeds we conservatively adopt $\eta_\textrm{\tiny{therm}}^\textrm{\tiny{max}} \equiv 1.0$ and $\eta_\textrm{\tiny{therm}}^\textrm{\tiny{min}} \equiv 0.1$ where, in this particular instance, we are concerned with only that proportion of the mechanical power injected into the ISM that accelerates the overall outflow.}).
This is somewhat larger than the $\sim$10\% thermalization efficiency determined for normal or quiescent starforming galaxies \citep[see, e.g.,][]{Strickland2009}, 
but this parameter is a function of local ISM conditions with, in particular, higher gas densities tending to imply larger radiative losses (and, hence, {\it smaller} thermalization values).
It is then somewhat surprising, {\it a priori}, that, in the molecular-gas rich milieu of the GC, the thermalization is rather high.
In fact, further consideration reveals the high thermalization is plausible: supernovae explode into the low-density cavity created by their massive progenitor star's wind and UV output \citep{Chevalier1999}. 
Moreover, in a starburst-like environment, one must consider the effect of previous SN which will tend to have rendered the local ISM highly porous -- riven with regions of high temperature, low density plasma -- through their own gas heating \citep{Heckman1990}. 
In any case, a thermalization efficiency of $>$40\% is in the range recently determined by \citet{Strickland2009} for the M82 starburst.

\subsubsection{Final estimate of SN rate}

We have, then, an estimate for the total supernova rate in the HESS field of 0.02--0.08/century, translating to a total kinetic power due to supernovae
of $L_{SN} = 1.3_{-0.6}^{+1.7 }\times 10^{40}$ erg/s (adopting the fiducial $10^{50}$ erg per supernova into relativistic particles decided above).
%\be
%L_{SN} = 1.3 \times 10^{40} \ \textrm{erg/s} \left(\frac{\nu_{SN}}{0.04/\textrm{century}}\right) \left(\frac{E_{SN}}{10^{51} \ \textrm{erg}}\right)
%\ee

\subsection{Supernova mass injection}

We have previously estimated \citep{Crocker2010b} the hot mass injected into a potential superwind -- that includes both supernova ejecta and the swept-up mass 
%(Heinz V{\" o}lk, private communication) 
-- on the basis of the study of \citet{Zirakashvili2006} of the nearby starburst NGC 253. 
On the basis of our previous study and
adopting the central value of the supernova rate, 0.04/century, 
one determines a mass injection rate due to supernovae in the HESS field of $\Delta\dot{M}_{SN} \simeq 0.01 \msun$/year.

\subsection{Winds off massive stars}

Though expected to make only a subdominant ($\sim$10\%)
contribution to the total mechanical power starbursts in general, 
stellar winds may supply up to half of the mass returned to  the ISM
\citep{Leitherer1992,Leitherer1995,Strickland2009}.
This expectation is thought to be borne-out in the GC \citep{Muno2004}.
\citet{Chevalier1992} originally calculated the wind power due to the massive (dominantly Wolf-Rayet) stars in the central pc of the Galaxy to be $\sim 5 \times 10^{37}$ erg/s for a mass loss rate from He {\sc i} emission line stars of $3 \times 10^{-4} \msun/$year and an assumed wind velocity of 700 km/s. 
Over the entire central 5 pc, the total mass loss rate from the Central stellar cluster is around $3 \times 10^{-3} \msun/$year \citep{Fryer2007} and that from the Arches and Quintuplet clusters have been determined to be $4 \times 10^{-4} \msun/$year and $3 \times 10^{-3} \msun$/year, respectively \citep{Lang2005}. 
Given that Paschen-$\alpha$ studies \citep{Wang2009} indicate that about half of the GC's massive stars are located {\it outside} these three clusters, we expect that stellar winds launch a total $\Delta\dot{M}_{wind} \simeq 0.01  
\msun/$year. 
Assuming a typical wind speed of 700 km/s, this implies a total kinetic power of $1.3 \times 10^{39}$, consistent with the fiducial 10\% of SN power expected.

\subsection{Total mass and power budgets}
\label{sctn_TotalMass}

Given the above, supernovae and (sub-dominantly) stellar winds inject, together, a mechanical power into the HESS region of $\sim 1.4 \times 10^{40}$ erg/s, uncertain by a factor of $\sim$2[\footnote{There are a number of other sources of mechanical power in the system which, while ranging from the virtually-assured to the speculative, are all rather poorly constrained empirically. One of the results of our global analysis of the system's particle astrophysics is that there is no evidence that power above and beyond that injected by supernovae (and stellar winds) is {\it required} in explaining the region's non-thermal emission at radio and TeV energies (the GeV spectrum may indeed, however, be dominated by the contribution of milli-second pulsars as discussed below). For completeness we list other potential sources of power below (\S\ref{section_OtherPower}).}].

The mass lost off massive stars is approximately equal to the mass returned to the ISM in supernovae and these components sum to give a total mass return of
\be
\Delta\dot{M}_{SN} + \Delta\dot{M}_{wind} \equiv \Delta\dot{M}_{tot} \simeq  0.02  - 0.03 \, 
\msun/ \textrm{year} \, .
\ee
This estimate is close to the range  $0.03  - 0.1 
\msun/ \textrm{year}$ estimated by \citet{Figer2004} to be lost into a thermally-driven wind (over a somewhat larger region). 
Also note that, on the basis of \S \ref{sctn_SFR}, the  mass injected by supernovae and stellar winds per star formation rate is $(0.02 - 0.03) \msun/$year$/(0.08 \msun/$year$) \ = \ (0.25 - 0.38)$ close to the range of $(0.19 - 0.31)$ arrived at by \citet{Strickland2009} in their detailed modelling of the starburst system M82.

\subsection{Other sources of power}
\label{section_OtherPower}

For completeness, we mention that other sources of power considered in the GC context include i) pulsars, in particular milli-second pulsars \citep{Wang2006} which are very difficult to detect in the GC environment because of pulse broadening of $\sim$100 $s$ at $\sim$ GHz in the turbulent plasma of the GC \citep{Cordes1997,Lazio2008} but which are generically expected to contribute around 10\% of the kinetic power produced by SN; ii) tidal disruption of stars which approach too close to the central SMBH, a process predicted to occur at a frequency $10^{-4} - 10^{-5}$ yr$^{-1}$, each event releasing $10^{51-52}$ erg \citep{Strubbe2009}, possibly leading to important radiative consequences \citep{Dogiel2010}; iii)
magnetic reconnection \citep{Tanuma1999}; iv) the gravitational energy of the region's giant molecular cloud tapped via the heating induced by the MHD waves excited  by these clouds' passage through the (hypothesised) very hot (and highly viscous) plasma \citep{Belmont2006}.

\section{Indications and Expected Characteristics of a GC Outflow}
\label{sectn_outflow}

There is considerable evidence -- reviewed here  -- 
for a rather fast GC outflow that would be expected to advect non-thermal particles from the system.
Most of our discussion below will focus on a
global GC outflow beyond the central region ($<$ 10 pc) surrounding
Sgr A$^*$. However, it should be noted that modelling of this environment
\citep[e.g.,][]{Chevalier1992,Ruffert1994}
has itself provided compelling
evidence for a substantial mass outflow from the inner few lightyears
of the Galaxy.

\subsection{Direct evidence for a GC outflow}

The strongest evidence in support of the existence of an outflow from the GC comes from observations of a $\sim1^\circ$ (or $\sim 140$ pc) tall and diameter $<130$ pc shell of emission rising north of the Galactic plane called the Galactic Centre Lobe. 
This structure was first identified in 10 GHz radio continuum emission \citep{Sofue1984} and has been interpreted as evidence for a previous episode of either starburst \citep{Chevalier1992,Bland-Hawthorn2003} or nuclear activity \citep{Kundt1987,Melia2001} or as due to magneto-dynamic effects \citep{Uchida1985}.
RC emission from the lobe's eastern part has HI absorption that clearly puts it in the GC region \citep{Lasenby1989}.

The coherent nature of the GC lobe was suggested by the discovery \citep{Bland-Hawthorn2003} at mid-infrared (MIR) wavelengths of filamentary structures coincident with the lobe's entire radio structure\footnote{Another suggestive association is between the Radio Arc -- the brightest and most dramatic of the GC's non-thermal radio filaments -- and the eastern extension of the lobe: \citet{Sofue1984,Yusef-Zadeh1988,Pohl1992}.}.
Modelling  \citep{Bland-Hawthorn2003} of the MIR emission from shock-heated dust pointed to outflow masses and kinetic energies of $5 \times 10^5 \msun$ and $> 10^{54}$ ergs, respectively, with an associated formation timescale of $\sim10^7$ years.

Most recently \citet{Law2008} and \citet{Law2010} have  demonstrated the coherent nature of the lobe on the basis of its very similar morphologies  in radio continuum, radio and optical recombination lines, and the MIR and \citet{Law2009} demonstrated that its ionized gas is of high metallicity, consistent with a GC origin.
This multi-wavelength analysis shows, in addition, that the entire structure has a similar spectral index at $\sim$GHz frequencies and that the structure is composed of multiple nested shells showing  from,  outside to inside, respectively, i) MIR emission from  entrained  dust and poly-cyclic aromatic hydrocarbons, ii)  synchrotron emission at RC wavelengths (diameter  $\sim$ 110 pc and with an equipartition \citep{Beck2005} field in the range 40 -- 100 $\mu$G), and iii) optical and recombination line emission from ionized material (which data also demonstrates the structure is located at the GC rather than being local).
Also of note is that the detailed analysis of \citet{Law2010} considerably revises downwards \citep[with respect to the estimate of][]{Bland-Hawthorn2003} the energetics of the structure to $\sim 5 \times 10^{52}$ erg 
and finds that the formation of the Lobe is consistent with currently-observed pressures and rates of
star-formation in the central few $\times 10$ pc  of the Galaxy\footnote{Note that even taking the previously-claimed  lower limit on the total energy associated with the outflow of $> 10^{54}$ erg, the mechanical power, $1.4\times 10^{40}$ erg/s, supplied by GC supernovae occurring at the rate we have claimed can potentially supply this over the claimed formation timescale $\sim10^7$ years.}.
%
%Finally, it is remarkable that the $\sim$ 150 pc extent of the bright region of 6.7 keV Fe line emission \citep{Yamauchi1990} -- 
%putatively tracing the 8 keV plasma inhabiting the inner GC -- is very similar to the that of the GC.

\subsection{Hints from external galaxies}

In the case of external star-burst systems and active galaxies the evidence for outflows is compelling \citep[e.g.,][]{Veilleux2005}.
Even for  less luminous, star-forming systems  akin the Milky Way (e.g., NGC 891)
the evidence for outflows from the central regions is strong  \citep{Dahlem1994,Breitschwerdt2008}.
For star formation surface densities exceeding $\dot{\Sigma}_*^{crit} \equiv 4 \times 10^{-2} \msun$ yr$^{-1}$ kpc$^{-2}$, empirical studies  \citep{Heckman1990,Strickland2009} suggest that 
supernovae and massive stars act in concert to drive a mass-loaded `super-wind' out of the starburst nucleus. 
\citep[Alternatively, though not independently given the existence of the Schmidt star-formation law, 
galaxies that exceed a critical 
surface density of gas of $\Sigma ^{crit} \equiv 0.05$ g cm$^{-2}$ are observed to drive  super-winds:][]{Heckman2000,Lacki2009}.

In general super-winds are comprised of multiple phases of varying density and velocity \citep{Strickland2009}. 
As revealed by observations of external, star-forming galaxies like M82, the `wind fluid' itself is a low-density and extremely hot  plasma with an initial temperature in the range $10^7 - 10^8$K. 

In any case, the areal star formation rate within the HESS region is $\dot{\Sigma}_* \sim 2  \msun$ yr$^{-1}$ kpc$^{-2}$, exceeding $\dot{\Sigma}_*^{crit}$ by almost two orders of magnitude.
From the point of view furnished by studies of other galaxies, then, the expectation is that the GC should be driving an outflow.

\subsubsection{Expected characteristics of a GC superwind}

As in external, star-forming galaxies, the mechanical power and mass injected collectively into the ISM by GC supernovae and massive stellar winds
might be expected to generate a `super' wind with the following characteristics:
\begin{enumerate}
\item
The asymptotic speed of such a wind \citep[which the outflow quickly reaches: e.g.,][]{Voelk1990} should scale as
 $v_{wind} \sim \sqrt{\dot{E}/\dot{M}}$ which evaluates to  800-1000 km/s $\dot{M}$ range nominated above (0.02-0.03 $\msun$/year and a {\it total} mechanical power of $1.4 \times 10^{40}$ erg/s and assuming thermalization of 100\%).
More accurately,  
the asymptotic speed of an outflow powered by the thermalized energy injected by supernovae and other sources in the disk of a starburst galaxy
can be rendered as \citep[e.g.,][]{Zirakashvili2006} :
\begin{eqnarray}
u_{\infty }%=\frac \gamma {\gamma -1}u(h)=
&\simeq& 1200 \ \mathrm{km\ s}^{-1} 
\sqrt{\eta_\textrm{\tiny{therm}}}
\left( \frac {  \dot{E}}{1.4 \times 10^{40}\mathrm{ \ erg\ s}^{-1}}\right) ^{1/2} \nn \\
&& \qquad  \times \left( \frac {\dot{M}} {0.03 \ M_\odot \ \mathrm{yr}^{-1}}\right) ^{- 1/2}, 
\label{eqn_ZV_vel}
\end{eqnarray}  
where $\dot{E}$ and $\dot{M}$ are the total energy and mass production rates and $\eta_\textrm{\tiny{therm}}$ parameterizes the thermalization efficiency of the overall superwind. 

\item The central temperature of the wind-launching region can be rendered as \citep{Strickland2009}:
\begin{eqnarray}
\label{eqn_Twind}
T_c &=& 0.4 \ \frac{\mu \ m_H \ \eta_\textrm{\tiny{therm}} \ \dot{E}}{k_B \ \beta_\textrm{\tiny{load}} \ \dot{M} } \nn \\
&\simeq &
2 \times 10^7 \ \textrm{K}
\
\frac{\eta_\textrm{\tiny{therm}}}{\beta_\textrm{\tiny{load}}}
\
\frac{\mu}{1/2}  \\
&& \qquad \times 
 \left( \frac {  \dot{E}}{1.4 \times 10^{40}\mathrm{ \ erg\ s}^{-1}}\right) \left( \frac {\dot{M}} {0.03 \ M_\odot \ \mathrm{yr}^{-1}}\right) ^{- 1} \, , \nn
\end{eqnarray}
where $\mu$ is the mean particle mass (in units of the proton mass) normalized to the expected value for a $H$ plasma and  $\beta_\textrm{\tiny{load}} \geq 1$
parameterizes the degree of mass-loading of the outflow.
%The $He$ is -- as noticed by  \citet{Belmont2005} -- gravitationally bound but probably constitutes the bulk of the wind fluid. 

\item The central pressure of the wind-launching region can be rendered as \citep{Strickland2009}:
\begin{eqnarray}
P_{\rm c} &=& 0.118 \, \frac{\eta_\textrm{\tiny{therm}}^{1/2} \, \beta_\textrm{\tiny{load}}^{1/2} \, \zeta \,  
  \dot{E}^{1/2} \, \dot{M}^{1/2}}{ k_B \ R_{\star}^{2}} \nn \\
&\simeq&   3.3 \times 10^6 \ {\rm K \ cm}^{-3} \  \eta_\textrm{\tiny{therm}}^{1/2} \, \beta_\textrm{\tiny{load}}^{1/2} \, \zeta \  \nn \\
&& \times
 \left( \frac {  \dot{E}}{1.4 \times 10^{40}\mathrm{ \ erg\ s}^{-1}}\right)^{1/2} \nn \\
 && \qquad \times \left( \frac {\dot{M}}
  {0.03 \ M_\odot \ \mathrm{yr}^{-1}}\right) ^{1/2}
\left( \frac {R_\star} {65 \ {\rm pc}}\right) ^{-2}  ,  
\label{eqn_Pc}
\end{eqnarray}
where $R_\star$ is the radius of the starburst region which we have normalized to the $\sim$65 pc radius of the GC Lobe \citep{Law2010} (and consistent with the extent of the bright X-ray plasma region) and the $\zeta \sim 1$ is the participation factor defined by \citet{Strickland2009} which parameterizes the fraction of total star formation that is actually involved in driving the superwind.

\item Lastly the central density of the wind-launching region can be rendered as \citep{Strickland2009}:
\begin{eqnarray}
\rho_{\rm c} &=& 0.296 \, \frac{ \zeta \, \beta_\textrm{\tiny{load}}^{3/2} \,  
  \dot{M}^{3/2}}{\eta_\textrm{\tiny{therm}}^{1/2} \,  
 \dot{E}^{1/2} 
  \, R_{\star}^{2}} \nn \\
  &\simeq&   2.5 \times 10^{-2} \ {\rm cm}^{-3} \  \eta_\textrm{\tiny{therm}}^{-1/2} \, \beta_\textrm{\tiny{load}}^{3/2} \, \zeta \  \nn \\
&& \times  \left( \frac {  \dot{E}}{1.4 \times 10^{40}\mathrm{ \ erg\ s}^{-1}}\right)^{-1/2} \nn \\
&& \qquad \times \left( \frac {\dot{M}} {0.03 \ M_\odot \ \mathrm{yr}^{-1}}\right) ^{3/2} 
\left( \frac {R_\star} {65 \ {\rm pc}}\right) ^{-2}
  .
\label{eqn_rhoc}
\end{eqnarray}

\end{enumerate}

\subsection{Additional evidence in support of the existence of GC outflow}
\label{sectn_WindExtraEvidence}

\subsubsection{Gamma-ray deficit in inner Galaxy}
\label{sectn_gRayDeficit}

A long-standing problem in CR studies \citep{Strong1996,Breitschwerdt2002} concerns why the $\gamma$-ray intensity across the inner Galaxy increases (relative to local) so much more slowly (for decreasing Galactic radius) than expected given the increase in areal density of plausible CR sources (and also given the growth in the density of ambient gas) towards the centre of the Galaxy.

This has a natural explanation \citep{Suchkov1993,Breitschwerdt2002} if particle escape times also {\it decline} in proportion to the increase in source density and this, in turn, is naturally explained if the {\it same} objects that are the ultimate sources of power for the non-thermal particles also serve to drive an increasingly fast wind to advect particles out of the Galactic plane
\citep[also cf.][]{Blitz1993}. Following this logic, GC CR density should roughly satisfy:
\be
n_{CR} \propto \frac{\dot{\Sigma}_*}{v_{wind}}
\ee
With $\sim$2\% of the Galaxy's star formation occurring in the inner 300 pc (diameter),  the over-abundance of CRs inside 300 pc with respect to the value in the Galactic disk should then be
\begin{eqnarray}
\frac{n^{GC}_{CR}}{n^{disk}_{CR}}  &\sim&  0.02 \ \left(\frac{15 \ \textrm{kpc}}{150 \ \textrm{pc}}\right)^2 \  \ \left( \frac{v^\textrm {\tiny{disk}}}{v_\textrm{wind}^\textrm{\tiny{GC}}} \right) \nn \\
&& \qquad \simeq 20 \left( \frac{v_\textrm{wind}^\textrm{\tiny{GC}}}{1000 \ \textrm{km/s}} \right)^{-1} \ , \nn
\end{eqnarray}
where the (diffusive) escape of CRs from the disk is parameterized via an effective $v^\textrm{\tiny{disk}} \equiv 10$ km/s \citep{Suchkov1993} and we normalize to a wind speed of order the gravitational escape speed \citep[as observed for external galaxies: e.g.,][]{Heesen2009}.

Implicitly here we find additional evidence that, in fact, a wind of a least a few 100 km/s is required in the GC: $\gamma$-ray studies at $\sim$GeV energies of the GC that failed to detect emission associated with some of the massive molecular clouds (Sagittarius B, C, or D) belonging to the CMZ \citep{Mayer-Hasselwander1998} require that the GC CR energy density be no more than a few times larger than the local value in the few GeV energy range. 
Note, however, that this constraint requires that CRs in this energy range sample all the mass of the ambient molecular clouds, including, in particular, the material in the cores of the clouds -- and theoretical  \citep{Dogiel1990,Gabici2007} and empirical studies \citep{Protheroe2008,Fujita2009}\footnote{Also see Jones, Crocker et al. 2010, `Australia Telescope Compact Array Radio Continuum 1384 and 2368 MHz Observations of Sgr B', submitted to the Astronomical Journal.} show this may not be true, potentially weakening this argument.

\subsubsection{Evidence of GC outflow on larger scales}

Across the electromagnetic spectrum, there is considerable evidence of GC outflows on scales much larger than the $\sim 1^\circ$ extension of the GC lobe, up to order tens of degrees.
This evidence includes:

\begin{enumerate}

\item  The North Polar Spur (NPS), a thermal X-ray/radio loop extending $\sim 80^\circ$ north of the Galactic plane first identified by \citet{Sofue1977,Sofue1984,Sofue2000} and interpreted by them as evidence for a $\gtrsim 10^{55}$ erg nuclear outburst $\sim 15$ Myr ago.
Note, however, that there are also strong arguments to suggest the NPS is a local ISM feature produced by the continuous effects of supernovae and stellar winds from the ScoÐCen association \citep[see, e.g.,][]{Miller2008}.

\item On scales of up to $\sim 25^\circ$ north of the GC, the {\it Galactic centre spur} identified \citep{Sofue1989} in 408 MHz \citep{Haslam1982} and 1408 MHz \citep{Reich1990} data. 

\item In general, ROSAT X-ray observations covering the central regions (inner few kpcs) of the Galaxy \citep{Sofue2000,Almy2000,Yao2007, Everett2008,Breitschwerdt2008} seem to indicate a nuclear or bulge outflow of plasma with a temperature few $\times 10^6$ K. The detailed analysis of \citet{Everett2008} suggested a Galactic wind towards the centre of the Galaxy asymptoting to 760 km/s in the halo and strongly suggested a role for CRs in helping to launch this wind (interestingly, however, these authors find that -- assuming constant ISM conditions fixed to match the local ones -- a wind cannot be launched inside $\sim$ 1.5 kpc due to the growth in the Galaxy's gravitational potential).

\item Radio polarimetry measurements \citep{Duncan1998} which show a plume of emission extending $\sim 15^\circ$ north of the Galactic plane, roughly centred on the GC, with weaker evidence for a mirroring southern structure.

\item From 21 cm observations, \citet{Lockman1984} determined that the inner Galaxy seems to be relatively depleted in H I, plausibly attributed to clearing of the neutral gas from the region $|z| > 500$ pc, $R \lesssim 3$ kpc by a nuclear outflow.

\item The infrared dust observations by \citet{Bland-Hawthorn2003} indicate an outflow structure on very large scales, approaching the size of the  North Polar Spur.

\item A number of studies find evidence (in UV absorption features of various background objects) of high-velocity material plausibly associated with a large-scale, Galactic nuclear outflow.
\citet{Keeney2006} find evidence (in sightlines towards two high-latitude active galactic nuclei) of material both escaping from and falling toward the centre of the Galaxy with a typical speed of 250 km/s, the latter feasibly indicating the fall-back of material in a bound Galactic wind or `fountain' (see below).
\citet{Zech2008} report the detection of high velocity clouds (HVC) of, significantly, super-solar metallicity along the sightline of star at a distance $\sim$ 7.5 kpc. 
In their view, the probable close proximity of the HVC material to the GC, its complex, multiphase structure, its high metallicity, and its large velocity all indicate an association with a Galactic nuclear outflow or fountain.

\item Though the connection to the GC is more circumstantial, it has long been recognized that a flow of hot gas from the Galactic disk to the halo is necessary to sustain the latter \citep[see, e.g.,][]{Shull1996}. 
Prime candidates for the requisite gas conduits are the so-called chimneys of gas that have been observed to form above 
the `super-bubbles' around OB
associations
in external galaxies and also, apparently, the Galactic disk \citep{Normandeau1996}.
Given the massive stellar population of the GC mirrors that found in such associations, it seems not unlikely that, by analogy,  
a chimney-like structure should form above the GC especially given the detection of super-bubble like structures associated with the
GC massive stellar clusters \citep[see][and references therein] {Rodriguez-Fernandez2001}.

\end{enumerate}

Finally, we note the existence of other less-direct indications of a GC outflow include the following:
\begin{enumerate}

%\item Gravitational escape velocity is $\sim$ 1000 km/s (Breitschwerdt1991: 900 km/s, Belmont and Tagger (??): 1200 km/s)...Breitschwerdt determine that outflows from starbursting systems with typically speed of escape velocity

%\item As already mentioned, were the ``very hot" ($\sim$ 8 keV) X-ray emitting plasma real, it would be gravitationally unbound and escape in an outflow of a few 100 km/s.

\item One explanation  \citep{Shore1999} of the non-thermal radio filaments (NTFs) posits that they are, in fact, akin to cometary plasma tails formed from the interaction of a hypothesised, large-scale, magnetized, $\sim$1000 km/s Galactic wind `draping' dense molecular clouds in the GC region.
%\footnote{As mentioned, some have interpreted the non-thermal radio filaments (NTFs) as evidence for a pervasive and regular poloidal field through the GC of $\sim$mG amplitude \citep{Morris1989}. 
%
%Even in the absence of a large-scale outflow, were cosmic rays to stream along this field structure at the Alfven velocity \citep{Kulsrud1969} and were the magnetic field amplitude to be in the suggested $\sim$mG range, cosmic rays would be escaping from the system with an energy-independent velocity of $\sim2200$ km/s \citep{Morris2007}.
%
%Note, however, that prevasive field amplitudes as high as 1 mG are in tension with our analysis.}.

\item \citet{Lu2008} report the detection of 10 X-ray filaments in the inner $\sim$ 15 pc around GC which they interpret as pulsar wind nebulae whose tails, mostly pointing outwards, again suggest a GC wind of velocity $\gtrsim 400$ km/s.
\citet{Markoff2010} summarizes evidence from Chandra for bi-polar lobes of hot gas extending over similar scales around Sgr A$^*$.
Mid-infrared observations of two cometary structures on arcsecond scales around Sgr A$^*$ \citep{Muzic2007} have recently been interpreted \citep{Muzic2010} as indicating the presence of a collimated outflow of at least a few 100 km/s driven either by the SMBH or stars in its vicinity.

\item Extreme Doppler-broadening of the 1.8 MeV line from $^{26}Al$ decay is evident in emission detected around the GC shows  and indicates the decaying nuclei have a (non-thermal) velocity of $\sim$500 km/s \citep{Naya1996}, consistent with a wind advecting the material out of the Galactic plane.

\end{enumerate}

\subsection{Stalling of GC outflow}
\label{section_Stalling}

For an adiabatic outflow powered by a wind, a sufficient condition for the flow to reach infinity is that the asymptotic wind velocity be faster than the gravitational escape speed. 
Given that the Galactic escape speed on relevant scales is $\sim$900 km/s, this condition will be satisfied over at least some of the parameter space we favour. 
However, for a radiative flow, the conditions on the wind velocity are more stringent if it is not to stall.
The case of a wind driven from a super stellar cluster located close to the GC (e.g., the Arches cluster)
has recently been explored by 
\citet{Rodriguez-Gonzalez2009}. 
These authors determine the stall height of an outflow as a function of metallicity and wind velocity and find over the relevant parameter space (500 km/s$<v_{wind}<800$ and 1 $<Z<10$) that heights between 1 and 50 kpc from the plane can be achieved \citep[note that radio recombination line observations have found an unusually low electron temperature that implies a high metal abundance for the ionized gas in the GC lobe:][]{Law2009}.

\section{Radio Data}
\label{sctn_RadioData}

Sources of radio data for the HESS and DNS radio spectra are listed in table \ref{GCsurveys}.

\begin{table}
	\centering	
       	 \begin{tabular}{|c|c|c|cc|c|l|}
	 \hline\hline
         $\nu$  & Telescope &  Beam & \small{Flux} & \small{density}  & Error & Ref. \\
 (GHz)  &&  ($'$) & HESS & DNS (Jy) && \\
	\hline
	.074$^*$ & VLA &  2 & 1,328 & 16,200 &6 \% & LaRosa \\
 &&&&&& et al.~(2005) \\
	.330 & GBT  & 39 & 2,545  &18,000$^\dag$ & 5\% &  LaRosa \\
 &&&&&& et al. (2005) \\
%	.330$^\ddag$ & VLA & 2 & 1,000 & - & \citet{LaRosa2000,LaRosa2005} \\

	%.408 & PKS & 64 & 51 & 800 mK & \cite{408MHz} \\

	1.408 & EBG  &  9.4 & 1,915 & 7,300 & 10\% &  Reich \\
 &&&&&& et al. (1990) \\
	2.417 & PKS  & 10.4 & 1,525 & 4,900 & 6\% &  Duncan \\
 &&&&&& et al. (1995) \\
	2.695 & EBG  &  4.3 & 1,460 & 4,400 & 10\% &  Reich \\
 &&&&&& et al. (1984) \\
%	5.00 & Green Bank  & ** & 1,186 &  --- & ** \% & \cite{Law2010} \\

	8.35$^{**}$ & GBT(ES)	      	     & 	  9.7    & 790 & 1186 &  5.6\% &  Langston \\
 &&&&&& et al. (2000) \\
%	8.57 & Green Bank 	      	     & 	  **    & 650 & --- &  **\% & \cite{Law2010}\\

	10.29 & NOB  & 2.9 & 710 & 1,400 & 7\% & Handa \\
 &&&&&& et al. (1987) \\
	14.35$^{**}$ & GBT(ES)    	     & 	   6.6   &  780 &1,200 & 8.3\% &Langston \\
 &&&&&& et al. (2000) \\
	\hline
        \end{tabular}
       	\caption{Observational data used to derive the spectrum for the HESS and DNS regions identified in the main text. Notes: $^*$At 74 MHz the large-angle  Galactic plane synchrotron background/foreground flux contribution is not measured (and hence not accounted for) due to the interferometric nature of the VLA observation. $^\dag$Total flux (measured with Greenbank single dish) supplied by Dr Crystal Brogan (private communication). 
	%$^\ddag$Flux attributable to discrete sources $<1^\circ.2$. 
	$^{**}$Greenbank Earth Station fluxes are lower limits.
	{\bf Key}: VLA = Very Large Array, GBT = Greenbank telescope, EBG = Effelsberg, PKS = Parkes, GBT(ES) = Greenbank Earth Station, NOB = Nobeyama.}
	\label{GCsurveys}
\end{table}

In addition to the radio data employed in the paper, in table
\ref{GCsurveys} we also list
8.35 and 14.35 GHz data \citep{Langston2000}. Unfortunately, the data at these frequencies
were processed with a median filtering algorithm that artificially attenuates the flux density of sources larger than $\sim$ 34$'$ so that these flux densities strictly represent lower limits. 
Given the strong concentration of the 8.35 GHz emission into structures on size scales less than $\sim 0.5^\circ$, however, we treat the 8.35 GHz flux as a datum to be fit by our $\chi^2$ procedure.
The 14.35 MHz datum
shows the expected \citep{Finkbeiner2004} up-break above 10 GHz due to the rising contribution from  spinning dust and is, therefore, not adopted as a constraint in our fitting. 

%\subsubsection{Radio flux contributed by discrete, small-scale sources}

%The discrete source contribution measured in this way is in fair agreement with the expectation for such given by equation 

\section{Detailed Justification of Assumption that Particle Transport is Energy-Independent}
\label{sectn_justification}

In our modelling we assume that the non-thermal particle transport timescale is energy-independent, a situation that is realised if, for instance, 
the removal of particles from the region is affected by their being advected away on a wind (rather than diffusing in an energy-dependent fashion by scattering on magnetic turbulence).
This implies that transport does not alter the slope of the injection spectrum (of either relativistic electrons or protons over the energy range where its associated timescale is the least). Also note that -- as the $pp$ cross-section is, to a fair approximation, also energy independent -- the steady-state, in situ proton spectrum has the same shape as the injection distribution (cf. the situation for the  steady-state, in situ electron spectrum where the different energy dependences of the various electron cooling processes certainly can introduce spectral features).

Although an energy-independent timescale is an assumption, we believe this to be physically well-motivated for a number of reasons
\begin{enumerate}

\item There is no spectral evidence for diffusion processing of the spectrum. The spectral index of the diffuse $\sim$ TeV emission is  2.2--2.3. 
It is firstly difficult to reconcile such a hard spectrum with any diffusion processing having occurred; in general, the least steepening, $\Delta \gamma = -1/3$ occurs for a Kolmogorov spectrum of turbulence. In this case, the injection spectral index must be very hard, 1.9--2.0. (Any other spectrum of turbulence would require an even harder injection spectrum.) Given that -- as we explain at length below -- we expect that supernova remnants accelerate the protons and primary electrons needed to explain the non-thermal emission detected from the region and that  the local CR spectrum (also due mostly to acceleration by SNRs in the Galactic disk) has a spectral index $\gtrsim 2.7$,  the most natural explanation is that both GC and local CRs have the same spectrum at injection (in the same class of astrophysical object) but, whereas the local CRs have been diffusion steepened, the GC CRs have not.
The only escape clauses to this argument would seem to be:
\begin{enumerate}
\item the system is out of steady-state with insufficient time having elapsed since an assumed, single injection event for the particle distribution to have been steepened \citep[cf.~the interpretation adopted by][that a single CR-injection event $\sim10^4$ year ago at the GC explains the observed diminution in the $\gamma$-ray to molecular column ratio beyond $|l| \sim 1^\circ$]{Aharonian2006}. 
As argued above, however, all the evidence is consistent with the system being in or close to steady-state. 

\item
the diffusive transport time,  $t_{dffsn} \equiv r^2/(4 D)$, is actually close to the rectilinear escape time $t_{rect} = r/c$ for a region size $r$. 
In this case, the diffusive escape time must become energy-independent so that the steady-state, in situ spectrum must flatten back to that at injection \citep{Aharonian2005}. 
An exploration of this scenario is beyond the scope of the present work but note, at least as far as the protons are concerned, this situation still realises an effectively energy-independent transport time as we have modelled.
For an application of this idea to the GeV/TeV point source coincident with Sgr A* see \citet{Chernyakova2010}.
 
\end{enumerate}
A related point is that, within the admittedly large statistical errors, the spectral index of the diffuse $\sim$ TeV emission is constant over the solid angle of the emission region. There is thus no evidence for steepening of the CR hadron distribution as the particles are transported -- and age -- away from their putative discrete sources.

\item The radio spectral information is also highly suggestive of energy-independent transport in the HESS field. According to the modelling results of \citet{Lisenfeld2000}, radio spectral indices $\alpha \lesssim 0.7$ (for $S_\nu \propto \nu^{-\alpha}$) indicate such and a power-law fit to the radio data from the HESS field for all radio data 1.4--10 GHz finds a spectral index of $\alpha \lesssim 0.54$, corresponding to an emitting electron population (for $F_e \propto E_e^{-\gamma}$) with spectral index $\gamma \simeq 2 (\alpha + 1) \lesssim 2.1$ (inclusion of the 74 MHz datum leads to an even harder spectra but this has been affected by free-free absorption).

\item Diffusive transport has been shown to be very slow in the GC by direct modelling of particle trajectories \citep{Wommer2008}. 

%\item Particle diffusive tranport {\it along} the Galactic plane in the GC can be expected to be particularly slow \citep{Morris2007} given that all the evidence points to the large scale field (outside molecular clouds) to have a poloidal geometry \citep{Chandran2000}

\item All the evidence from starbursts -- to which we have argued the GC is closely analogous -- is that CR transport is overwhelmingly advective \citep{Suchkov1993,Dogiel2009}.
Theoretically, this is expected in light of the fact that CR scattering produced by self-excited Alfven waves should be very effective in starbursts,  leading to marked reductions in the predicted diffusion coefficient \citep{Ptuskin1997,Zirakashvili2006}.
On the empirical side,
in the well-known star-burst system M82, for instance, cosmic ray escape is probably dominated by advection out of the gas midplane in a starburst-driven wind \citep{Thompson2007,Klein1988,Seaquist1991}.
\end{enumerate}

\end{document}